\documentclass[sigconf]{acmart}
\AtBeginDocument{%
  }

\setcopyright{none}
\copyrightyear{}
\acmYear{}
\acmDOI{}
\acmConference[Conference acronym 'XX]{Make sure to enter the correct
  conference title from your rights confirmation emai}{June 03--05,
  2018}{Woodstock, NY}
\acmISBN{}
\renewcommand\footnotetextcopyrightpermission[1]{}

\usepackage{multirow}
\usepackage{enumitem}
\usepackage[most]{tcolorbox}

\newcommand\eg{{\it e.g.}}

\begin{document}

\title{GraphRAG-Induced Dual Knowledge Structure Graphs for Personalized Learning Path Recommendation}

\author{Xinghe Cheng}
\orcid{0000-0001-9432-5794}
\affiliation{
  \institution{Jinan University}
  \city{Guangzhou}
  \country{China}
}
\email{jnuchengxh@hotmail.com}

\author{Zihan Zhang}
\orcid{0009-0004-7079-8705}
\affiliation{
  \institution{Jinan University}
  \city{Guangzhou}
  \country{China}}
  \email{zzihan98@163.com}

\author{Jiapu Wang}
\orcid{0000-0001-7639-5289}
\affiliation{
  \institution{Beijing University of Technology}
  \city{Beijing}
  \country{China}}
  \email{jiapuwang9@gmail.com}
  
\author{Liangda Fang}
\orcid{0000-0002-6435-6570}
\authornote{Corresponding Author.}
\affiliation{
  \institution{Jinan University\\ Pazhou Lab}
  \city{Guangzhou}
  \country{China}}
  \email{fangld@jnu.edu.cn}
  
\author{Chaobo He}
\orcid{0000-0002-6651-1175}
\affiliation{
 \institution{South China Normal University}
  \city{Guangzhou}
  \country{China}}
  \email{hechaobo@foxmail.com}

\author{Quanlong Guan}
\orcid{0000-0001-6911-3853}
\authornote{Corresponding Author.}
\affiliation{
  \institution{Jinan University}
  \city{Guangzhou}
  \country{China}}
  \email{gql@jnu.edu.cn}

\author{Shirui Pan}
\orcid{0000-0003-0794-527X}
\affiliation{
  \institution{Griffith University}
  \city{Gold Coast}
  \country{Australia}}
  \email{s.pan@griffith.edu.au}

\author{Weiqi Luo}
\orcid{0000-0001-5605-7397}
\affiliation{
  \institution{Jinan University}
  \city{Guangzhou}
  \country{China}}
  \email{lwq@jnu.edu.cn}

\renewcommand{\shortauthors}{Xinghe Cheng et al.}

\begin{abstract}
Learning path recommendation seeks to provide learners with a structured sequence of learning items (\eg, knowledge concepts or exercises) to optimize their learning efficiency.
Despite significant efforts in this area, most existing methods primarily rely on prerequisite relationships, which present two major limitations:
1) Requiring prerequisite relationships between knowledge concepts, which are difficult to obtain due to the cost of expert annotation, hindering the application of current learning path recommendation methods.
2) Relying on a single, sequentially dependent knowledge structure based on prerequisite relationships implies that difficulties at any stage can cause learning blockages, which in turn disrupt subsequent learning processes.
To address these challenges, we propose a novel approach, GraphRAG-Induced Dual \textbf{Know}ledge Structure Graphs for Personalized \textbf{L}earning \textbf{P}ath Recommendation (KnowLP), which enhances learning path recommendations by incorporating both prerequisite and similarity relationships between knowledge concepts.
Specifically, we introduce a knowledge concept structure graph generation module EDU-GraphRAG that adaptively constructs knowledge concept structure graphs for different educational datasets, significantly improving the generalizability of learning path recommendation methods.
We then propose a Discrimination Learning-driven Reinforcement Learning (DLRL) module, which mitigates the issue of blocked learning paths, further enhancing the efficacy of learning path recommendations.
Finally, we conduct extensive experiments on three benchmark datasets, demonstrating that our method not only achieves state-of-the-art performance but also provides interpretable reasoning for the recommended learning paths.
\end{abstract}

\ccsdesc[500]{Applied computing~E-learning}
\ccsdesc[500]{Information systems~Recommender systems}

\keywords{Learning Path Recommendation, Discrimination Learning, Reinforcement Learning}

\maketitle

\section{Introduction}
\begin{figure*}
  \centering
  \includegraphics[width=\linewidth]{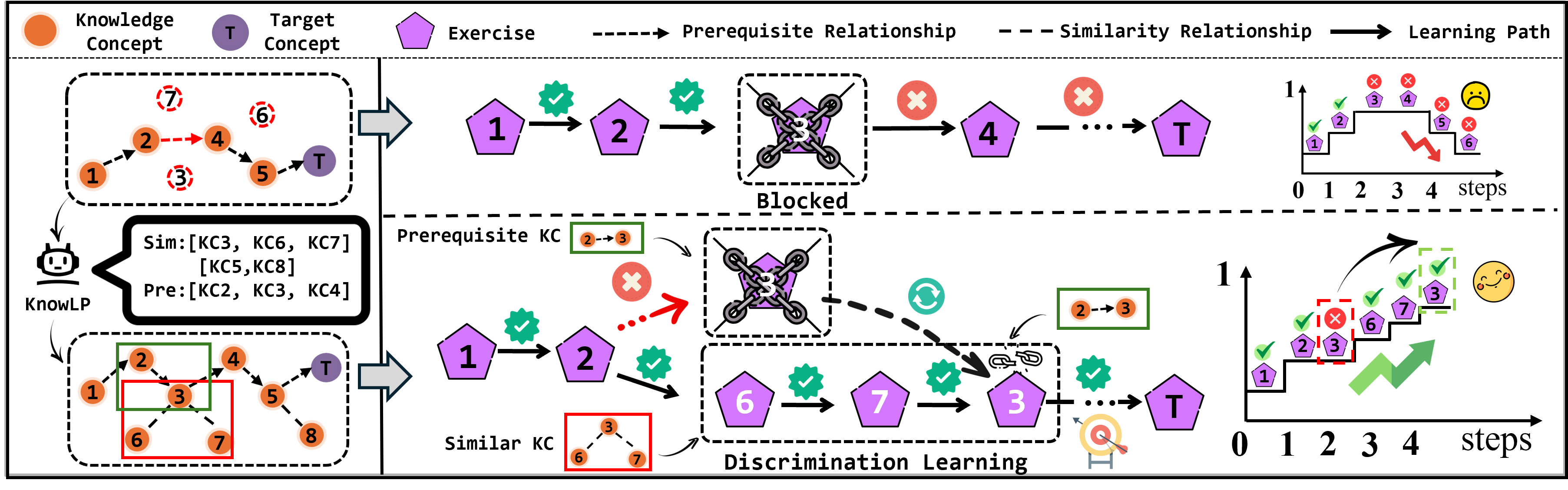}
  \caption{The comparison of traditional methods (top) and our proposed KnowLP (bottom) is depicted. The top denotes traditional KC prerequisite relationship-based methods where confusion between similar KCs such as $\text{KC}_3$, $\text{KC}_6$, and $\text{KC}_7$ can hinder students' mastery of $\text{KC}_3$ and impede their understanding of subsequent KCs. The bottom illustrates how KnowLP enables students to more effectively differentiate between these similar KCs, thereby improving learning outcomes.} 
  \label{fig:case}
\end{figure*}
With the rapid advancement of online education platforms, there is a growing demand for personalized learning experiences tailored to individual learners. In this context, Learning Path Recommendation (LPR) has emerged as a critical task, aiming to construct coherent and customized sequences of learning resources that align with specific educational goals \cite{WanLY2023,ZhaLZ2021,Che2009,ShiWX2020}. By organizing content in a pedagogically meaningful order, LPR facilitates systematic and progressive knowledge acquisition, thereby enhancing learning efficiency and effectiveness \cite{NabMP2017, GuaQL2025, CheXH2025}.

\looseness=-1
Prior research in education has demonstrated that the structure of Knowledge Concepts (KCs) plays a pivotal role in shaping effective learning path recommendations \cite{LiuTL2019}. Existing LPR approaches that incorporate KC structure can be broadly categorized into two paradigms: (i) correlation-based methods \cite{CheSX2023}, which infer personalized learning paths by exploiting implicit semantic or statistical associations among KCs; and (ii) prerequisite-based methods \cite{LiXY2024,ZhaSX2024}, which utilize explicit prerequisite relationships to guide the sequential organization of learning content. Among these, prerequisite-based approaches have gained particular attention due to their strong pedagogical grounding and consistency with human cognitive development, offering more interpretable and cognitively aligned learning trajectories.

\looseness=-1
Prerequisite relationships among KCs serve as a foundational scaffold for learning path recommendation, enabling stepwise progression toward learning objectives. However, the construction of such relationships often relies on costly expert annotations, thereby limiting the applicability of prerequisite-based methods \cite{ZhaSX2024}. Even when prerequisite graphs are available, they frequently suffer from sparsity and noise due to missing concepts or inaccurate dependencies. Recent advances in Large Language Models (LLMs) offer a compelling alternative by inferring prerequisite structures from natural language. Yet, their effectiveness remains constrained by the limited availability of high-quality textual descriptions in practice and the inherent risk of generating hallucinated or misleading links.

\looseness=-1
Moreover, prerequisite relationships often enforce a rigid sequential structure, where difficulties in mastering a single concept can cascade into broader disruptions along the learning path \cite{PanSM2024}. This structural inflexibility underscores the need for adaptive mechanisms that enable learners to recover when encountering conceptual obstacles. According to Gagné’s learning hierarchy \cite{Chi2022}, discrimination learning is essential for distinguishing between closely related knowledge concepts (KCs) \cite{Roh2012}. Integrating similar KCs into the learning process provides a contrastive scaffold, allowing learners to resolve ambiguities and strengthen conceptual differentiation. As shown in Figure~\ref{fig:case}, this approach improves learning resilience and promotes smoother progression, even in the presence of incomplete or uncertain prerequisite structures.

\looseness=-1
To address the aforementioned challenges, we propose KnowLP (\underline{Know}ledge structure graph retrieval-augmented generation for \underline{L}earning \underline{P}ath recommendation), a novel framework that jointly incorporates prerequisite and similarity relationships between KCs to construct more effective and adaptive learning paths.
Specifically, we introduce a KC structure graph generation module EDU-GraphRAG that adaptively constructs task-specific KC graphs. This module leverages TextGrad \cite{YukBB2025} to reduce LLM hallucinations, enabling the generation of accurate and densely connected KC structures tailored to diverse educational datasets.
Based on the generated graphs, we design a Discrimination Learning-driven Reinforcement Learning (DLRL) module composed of two cooperative agents: a prerequisite agent and a similarity agent. The prerequisite agent decides whether the learning trajectory should adhere to prerequisite dependencies, while the similarity agent detects conceptually confusable KCs in blocked learning scenarios and integrates them into the learning path to promote discrimination learning.

\looseness=-1
Our main contributions are summarized as follows:
\begin{itemize}[left=0cm]
    \item We propose KnowLP, a novel framework for personalized learning path recommendation, which adaptively constructs KC relationships from the target data and effectively generates learning paths through the collaboration of prerequisite, similarity, and difficulty agents.
    \item We propose an EDU-GraphRAG module that adaptively generates prerequisite and similarity relationship graphs for KCs on different datasets, with the goal of completing missing or incomplete knowledge structure graphs.
    \item We develop a DLRL module that introduces similarity relationships as a fallback when prerequisite relationships become ineffective, thereby alleviating the blocked phenomena.
    \item We conduct extensive experiments on three educational datasets, demonstrating that our method not only achieves state-of-the-art performance but also generates more effective longer learning paths.
\end{itemize}

\section{Related Work}
\subsection{Learning Path Recommendation}
Learning path recommendation is a key task in online education, focused on crafting personalized, structured learning paths for different learners to optimize their learning outcomes.
Existing methods can be categorized into two types according to the approach of KC structures considered: (1) KC correlation-based: learning paths are generated by inferring the latent correlations KCs. (2) KC prerequisite relationship-based: learning paths are generated based on the prerequisite relationships between KCs.

\looseness=-1
For branches of KC correlation-based approaches, many works have been proposed using different techniques, such as decision tree classifiers \cite{LinYH2013}, matrix factorization \cite{NabGG2020}, bayes theorem \cite{XuWC2012} etc.
One of the key contributions is the work by \citet{CheSX2023}, who proposed the SRC framework.
This framework models the complex relationships and interactions among KCs, enabling the generation of complete learning paths.
Due to its ability to better account for the sequential dependencies between KCs, KC prerequisite relationships-based methods are rapidly gaining attention.
This branch frequently treats learning path recommendation as a sequential decision-making problem, which results in the frequent application of advanced reinforcement learning techniques \cite{IntKT2020, CaiXY2018}.
For example, \citet{LiuTL2019} proposed a framework for adaptive learning utilizing cognitive structure, known as CSEAL, which integrates learners' knowledge levels with the KC prerequisite relations of learning materials to provide personalized learning paths.
\citet{LiXY2023} proposed a Graph Enhanced Hierarchical Reinforcement Learning (GEHRL) framework, which incorporates a graph-based candidate selector to constrain the action space of the low-level agent and employs a test-based internal reward mechanism to alleviate the issue of sparse external rewards.
\citet{ZhaSX2024} proposed a difficulty driven hierarchical reinforcement learning (DLPR) framework, which constructed a hierarchical graph of learning and practice items to capture the difficulty of items and their higher-order relations.
Despite their success, due to incomplete KC prerequisite relationships and neglect of KC similarity relationships, leading to inappropriate learning path recommendations.

\subsection{Retrieval-Augmented Generation}
Retrieval-Augmented Generation (RAG) is a well-established method for answering user queries across large datasets\cite{RamLD2023,LasHH2020,GooSD2020}.
It is specifically designed for scenarios where the answers are localized within specific text regions, and retrieving these regions provides adequate grounding for the subsequent generation task \cite{LewPP2020}.
When applying large language models (LLMs), RAG first retrieves relevant information from external data sources and then incorporates this information into the context window of LLM along with the original query \cite{RamLD2023, WanJP2024}.
Na\"{i}ve RAG approaches achieve this by first converting documents into text, then splitting the text into chunks, and finally embedding these chunks into a vector space where similar positions correspond to similar semantics \cite{GaoXG2023}.
In contrast to other RAG approaches, \citet{EdgTC2024} proposed a Graph RAG method that leverages global summarization of an LLM-derived knowledge graph, highlighting a previously unexplored aspect of graph structures in this context.
However, Graph RAG methods typically require sufficiently accurate documentation for effective implementation.
To obtain accurate documentation for KC descriptions, we integrated TextGrad \cite{YukBB2024} and developed a KC structure graph generation module.
This module mitigates the hallucination problem in LLMs and generates structural graphs of KCs based on their names, which not only adapts to generate structural graphs across different datasets containing KC names but also effectively supports most existing prerequisite-based learning path recommendation approaches.

\begin{figure*}
  \centering
  \includegraphics[width=\linewidth]{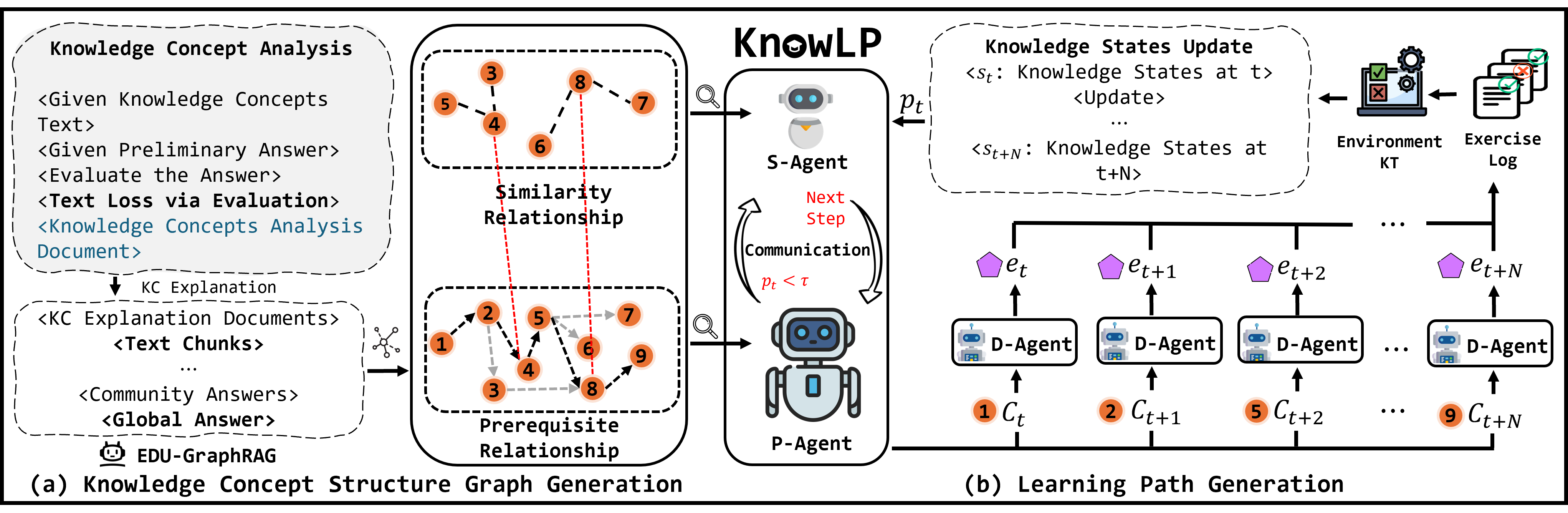}
  \caption{KnowLP Framework Overview. The KC structure graph generation module on the left begins by inputting KC names into TextGrad, which then generates textual descriptions. These descriptions are subsequently used by the EDU-GraphRAG to construct KC structure graphs. On the right, the learning path generation module initiates with the P-Agent and S-Agent filtering KCs. Following this, the D-Agent selects exercises of appropriate difficulty. The module then utilizes an environment KT to dynamically update the knowledge state in each session. Finally, the module generates the learning path step by step.}
  \label{KnowLP}
\end{figure*}

\section{Preliminaries}
In this section, we discuss the formulation of learning path recommendations and the definition of GraphRAG.

\paragraph{Problem Definition} The primary objective of this paper is to address the knowledge concept (KC) structure-based, goal-oriented learning path recommendation problem.
Suppose we have a set of exercises $E=\{e_1,e_2,\cdots,e_M\}$ and a set of KCs $C=\{c_1,c_2,\cdots,c_N\}$.
For each learner, learning goals are typically defined as $ G = \{g_1, g_2, \cdots\}$, where $G\subseteq C$.  
Each historical record $h=(e, \text{score})$ consists of an exercise $e \in E$ and the corresponding performance score.
Given a sequence of such records $H=\{h_1,h_2,\cdots, h_i\}$, where each $h_j=(e_j, \text{score}_j)$, and guided by the learning goal $g \in G$, a learning path $P=\{\tilde{e}_1,\tilde{e}_2,\cdots,\tilde{e}_k\}$ is recommended, where $\tilde{e}_j \in E$ denotes a recommended exercise.

\looseness=-1
To evaluate the effectiveness of the recommended learning path, we calculate the improvement in the learner’s feedback scores between the beginning and the end of the session.
The effectiveness of the learning path recommendation, denoted as $E_p$ is computed as follows:
    \begin{equation} \label{equation:learning score}
        \begin{aligned}
             E_p=\frac{E_{\mathrm{end}}-E_{\mathrm{start}}}{E_{\mathrm{sup}}-E_{\mathrm{start}}},
        \end{aligned}
    \end{equation}
where $E_{\mathrm{sup}}$ represents the maximum achievable score for the learning goals, which corresponds to the total number of learning objectives.
Our objective is to maximize the $E_p$ by provide an effective learning path.

\paragraph{Graph Retrieval-Augmented Generation}
Given a natural language query $ Q $ and a source document $D$, GraphRAG~\cite{EdgTC2024} aims to transform unstructured textual content into a structured graph representation $\phi(D)$, which facilitates accurate and explainable answer generation.
To achieve this, the extracted semantic elements, such as entities and relations, are passed to a graph constructor $\mathcal{G}(\cdot, \cdot)$, which aggregates and deduplicates them into coherent local subgraphs.
At query time, the model generates the final output $A$ by conditioning an LLM on the query $ Q $ and the structured graph index $\phi(D)$:
\begin{equation} \label{equation:graphrag_2}
    \begin{aligned}
       A = \mathrm{LLM}(Q, \phi(D))=\mathrm{LLM}(Q, \bigcup_{i=1}^{n} \mathcal{G} \left( \mathcal{E}, \mathcal{R} \right)),
    \end{aligned}
\end{equation}
where $\mathcal{E}$ denotes the set of extracted entities and $\mathcal{R}$ denotes the set of corresponding relationships among them. 

\section{Methodology}
In this section, we introduce our approach, KnowLP, shown in Figure~\ref{KnowLP}, which consists of two main components: Knowledge Structure Graph Generation and Learning Path Generation.
Knowledge Structure Graph Generation is responsible for constructing high-quality knowledge concept (KC) structure graphs by adaptively discovering both prerequisite and similarity relationships on different datasets.
Learning Path Generation, in turn, improves the effectiveness of longer learning paths by dynamically incorporating similarity relationships between knowledge concepts.

\subsection{Knowledge Structure Graph Generation}
In this subsection, we present a concise overview of the KC Structure Graph Generation module. We begin by leveraging TextGrad \cite{YukBB2025} to analyze the names of KCs, enabling more comprehensive retrieval of explanatory texts. Based on these texts, we design an EDU-GraphRAG framework that automatically constructs KC structure graphs, which serve as a crucial foundation for learning path recommendation.
\begin{tcolorbox}[colback=blue!5!white, colframe=black!75!black, title=Prompt for KC Explanation Generation]
\textbf{LLM:} (``You are an expert in the filed of education and are responsible for the detailed interpretation of knowledge concepts. The following are the knowledge concepts. Each knowledge concept is separated by `,':{knowledge concepts}.\\
\textbf{Question:} Use the knowledge in education to analyze the meaning of the serial number is {begin} to {end}. Knowledge concepts are: {knowledge concepts[begin:end+1]}, as well as detailed analysis the relationship between each of these knowledge concepts and other knowledge concepts.\\
\textbf{Output Format:} [knowledge concept]:[The analysis of knowledge concept.]\\
\textbf{Evaluation Instruction:} Make the generated analysis smarter, more logical and accurate, more specific and discriminative rather than vague, and avoid ambiguity. Ensure that the analysis of knowledge concepts is correct.")
\end{tcolorbox}

\subsubsection{Knowledge Concept Explanation Generation}
To improve the reliability of KC explanations, we are inspired by the TextGrad framework \cite{YukBB2025} to develop an iterative generation–evaluation– refinement procedure tailored to our task.
We formulate the KC explanation refinement as a multi-turn optimization process. The initial explanation $\mathcal{T}^{(0)}$ is generated using a predefined prompt $P_{\mathrm{gen}}$. At each iteration $t$, the explanation $\mathcal{T}^{(t)}$ is evaluated by a secondary LLM with evaluation prompt $P_{\mathrm{eval}}$, producing feedback $\nabla_{\mathrm{LLM}}^{(t)}$ via an additional feedback prompt $P_{\mathrm{feedback}}$. The explanation is then refined with a rewriting prompt $P_{\mathrm{rewrite}}$ as:
\begin{equation}
\mathcal{T}^{(t+1)} = \mathrm{LLM}\left(P_{\mathrm{rewrite}}, \mathcal{T}^{(t)}, \nabla_{\mathrm{LLM}}^{(t)}\right),
\end{equation}
\begin{equation}
\nabla_{\mathrm{LLM}}^{(t)} = \mathrm{LLM}\left(P_{\mathrm{feedback}}, \mathrm{LLM}\left(P_{\mathrm{eval}}, c, \mathcal{T}^{(t)}\right)\right).
\end{equation}

\looseness=-1
This iterative process continues until convergence or a predefined maximum step $T$ is reached.

\looseness=-1
In this process, the knowledge concept name serves as the input. Through prompt engineering, the LLM is guided to act as a domain expert in education and generate an initial explanation.
This explanation is then evaluated by another LLM to identify potential flaws, and iteratively refined based on the feedback until it reaches the desired level of accuracy and reliability.
The finalized explanation is designated as the source document $D_c$ for the knowledge concept, to be consumed in the next stage of our framework.
For instance, the objective function for the KC analysis is shown in the \textit{Prompt for KC Explanation Generation} prompt box.

\subsubsection{EDU-GraphRAG}
Given a collection of $N$ concept-level explanation documents $\{D_{c_1}, D_{c_2}, \dots, D_{c_N}\}$, where each $D_{c_i}$ provides a contextualized explanation of a distinct knowledge concept $c_i$, we construct a unified document $D_c$ by concatenating all these explanations in sequence. 
Next, the document $ D_c $ is segmented into a sequence of text chunks using a parameterized segmentation function $ \sigma(D_c; \theta) = \{ \sigma_1(D_c; \theta), \dots, \sigma_n(D_c; \theta) \}$.
For each chunk, an LLM abstractively extracts entities and relationships, capturing both explicit and implicit semantics in the text. These extracted components are passed to a graph constructor, which aggregates and deduplicates them into coherent local subgraphs. The global document-level graph index is obtained by unifying these subgraphs:
\begin{equation} \label{equation:graphrag_1}
    \begin{aligned}
        \phi(D_c) = \bigcup_{i=1}^{n} \mathcal{G} \left( \mathcal{E}(\sigma_i(D_c; \theta)), \mathcal{R}(\sigma_i(D_c; \theta)) \right).
    \end{aligned}
\end{equation}

At query time, given a query $Q$, EDU-GraphRAG conditions an LLM on the structured graph index $\phi(D_c)$ to generate the final output $A_\mathcal{G}$, which in our case corresponds to a constructed knowledge structure graph $\mathcal{G}$.
\begin{equation}
A_\mathcal{G} = \mathrm{LLM}(Q, \phi(D_c)),
\end{equation}
where $A_\mathcal{G} = (\mathcal{C}, \mathcal{S})$ denotes the induced knowledge structure graph comprising prerequisite and similarity relations among knowledge concepts, with $\mathcal{C}$ as the set of knowledge concepts and $\mathcal{S}$ as the set of semantic edges.

\subsection{Learning Path Generation}
In this subsection, we introduce the \textit{Discrimination Learning-driven Reinforcement Learning (DLRL)} module, which dynamically generates personalized learning paths while addressing learning stagnation. To model the learner’s evolving knowledge state, we adopt the \textit{Difficulty Matching Knowledge Tracing (DIMKT)}~\cite{SheHL2022} as the environment. The decision-making process is guided by a \textit{multi-agent framework}. Specifically, \textit{Prerequisite Agent (P-Agent)} and  \textit{Similarity Agent (S-Agent)} jointly track learning progress and select the next KC based on either prerequisite or similarity relationships, and \textit{Difficulty Agent (D-Agent)} selects exercises matched to the learner’s current ability level, ensuring a smooth progression in task difficulty.

\subsubsection{Knowledge Tracing}
The dynamic changes in a learner's knowledge mastery level during the learning process serve as a reflection of their learning outcomes.
By leveraging these insights to devise subsequent recommendation strategies, it is possible to provide a learning path that aligns with the learner's abilities and progress.
DIMKT is particularly advantageous as it can predict the impact of exercise difficulty on learners, thereby enhancing the accuracy of recommending exercises based on their difficulty level.

\looseness=-1
In this process, we employ the same HGNN \cite{YanYP2023, ZhaZW2023, ZheCP2023} embedding technique as utilized in DLPR \cite{ZhaSX2024} to embed both exercises and difficulty levels as inputs to DIMKT.
This method effectively integrates exercise features with difficulty vectors, ensuring that the predicted knowledge states comprehensively account for the interrelations between exercises and their corresponding difficulty levels.

\subsubsection{Prerequisite Agent}
To ensure that the generated learning path follows a pedagogically appropriate order for the learner, we develop the P-agent to select knowledge concepts sequentially based on their prerequisite relationships.

\looseness=-1
\textit{a) State encoder:}
At each step, the predicted knowledge state is based on the learner's learning objectives and historical learning records.
Thus, at the end of the $t-1$ step, the state of the P-agent includes the learner's knowledge state $\mathbf{h}_{t-1}$ and learning objectives $\mathbf{G}$.
Specifically, we use a one-hot encoding scheme to represent the learning objective $\mathbf{G}=\{0,1\}^M$, where $M$ is the total number of knowledge concepts.
The index corresponding to the learning objective is set to $1$, while all other indices are set to $0$.
At step $t$, the state of the P-agent $\mathbf{s}_t$ is represented as follows:
\begin{equation} \label{equation:relation agent}
     \begin{aligned}
        \mathbf{s}_t=\mathbf{h}_{t-1}\oplus \mathbf{G}.
        \end{aligned}
    \end{equation}

\looseness=-1
\textit{b) Policy:}
We adopt Proximal Policy Optimization (PPO) \cite{SchWD2017} as the model for the prerequisite relation agent, as it is one of the most performant reinforcement learning models currently available.
PPO is particularly effective at identifying the optimal sequence of knowledge concepts to achieve the learning goal.
Specifically, PPO consists of two primary components: a policy network (Actor) $\pi ( \mathbf{G}\left| \mathbf{s}_t;\theta \right.)$ and a value network (Critic) $V( \mathbf{s}_t;\phi)$, where $\theta$ and $\phi$ represent the respective network parameters.

\looseness=-1
The goal of the policy network is to generate a probability distribution over the knowledge concepts in the dataset. The state of the P-agent is fed into the policy network, which outputs a corresponding probability distribution over the KCs.
\begin{equation} \label{equation:policy}
     \begin{aligned}
        \mathbf{G} \sim \pi ( \mathbf{G}\left| \mathbf{s}_t;\theta \right. ) =\mathrm{Softmax}( \mathrm{Linear}(\mathbf{s}_t)),
        \end{aligned}
    \end{equation}
where $\mathrm{Linear}$ is the fully connected layer.
The value network takes the state of the relation agent as input and evaluates the reward associated with that state.
\begin{equation} \label{equation:reward}
     \begin{aligned}
        V\left( \mathbf{s}_t;\phi \right) =\mathrm{Linear}\left( \mathbf{s}_t \right).
        \end{aligned}
    \end{equation}

\looseness=-1
During each training step, we optimize PPO using a loss function similar to GEHRL \cite{CheSX2023}. The value network is trained using the mean squared error (MSE) loss function with gradient descent.
\begin{equation} \label{equation:loss function}
     \begin{aligned}
        L\left( \phi \right) =\mathbb{E} \left( \left\| \sum_{i=0}^{T-t}{\gamma ^iRe_{t+i}-V\left( \mathbf{s}_t;\phi \right)} \right\| \right) ^2,
        \end{aligned}
    \end{equation}
where $Re_t$ represents the reward at step $t$, the computation of which will be detailed later, and $\gamma$ denotes the discount factor.
Based on existing reinforcement learning optimization algorithms, we optimized the policy network using PPO-clip \cite{CheSX2023}.
\begin{equation} \label{equation:PPO-clip-1}
     \begin{aligned}
        \hat{A}_t=-V\left( \mathbf{s}_t;\phi \right) +Re_t+\cdots +\gamma ^{T-t}V\left( \mathbf{s}_T;\phi \right),
        \end{aligned}
    \end{equation}
\begin{equation} \label{equation:PPO-clip-2}
     \begin{aligned}
        L( \theta ) =\hat{\mathbb{E}}_t\left[ \mathrm{min}\left( r_t\left( \theta \right) \hat{A}_t,\mathrm{clip}\left( r_t\left( \theta \right) ,1-\epsilon ,1+\epsilon \right) \hat{A}_t \right) \right],
        \end{aligned}
    \end{equation}
where $r_t(\theta)$ represents the probability ratio $\frac{\pi ( \left. G \right|s_t;\theta)}{\pi \left( \left. G \right|s_t;\theta _{old} \right)}$, $\epsilon=0.2$ is a hyperparameter, and $\mathrm{clip}( r_t( \theta ) ,1-\epsilon ,1+\epsilon ) \hat{A}_t$ modifies the surrogate objective by clipping the probability ratio, which removes the incentive for moving $r_t$ outside of the interval $[1-\epsilon, 1+\epsilon]$.

\looseness=-1
\textit{c) Reward:}
To maximize the improvement in learners' mastery of the learning objectives, we calculate the corresponding reward value after selecting knowledge concepts at each step.
\begin{equation} \label{equation:MKS}
     \begin{aligned}
        Re_t=\begin{cases}
	E_p,\,\, \text{if t is the last learning stage}\\
	0,\,\,\,\,   \mathrm{otherwise.}\\
\end{cases}
        \end{aligned}
    \end{equation}

\begin{table*}[ht]
\centering
\caption{Performance comparison for learning path recommendation methods. Existing state-of-the-art results are underlined and the best results are bold. Our KnowLP is compared with the SOTA DLPR and $*$ indicates a $p-\text{value}<0.05$ in the t-test.}
\begin{tabular}{lcccccccccc}
\toprule
Dataset     & Steps & KNN    & GRU4Rec & RL-Tutor    & Actor-Critic & CSEAL  & SRC & GEHRL  & DLPR  & KnowLP     \\ \midrule
\multirow{3}{*}{Junyi} 
            & 5     & 0.1296 & 0.1504  & 0.1601 & 0.1891   & 0.1964 & 0.1874  & 0.1762 & \underline{0.2086} & \textbf{0.2406}$^*$ \\
            & 10    & 0.1485 & 0.2129  & 0.1923 & 0.2051   & 0.1779 & 0.2075  & 0.2222 & \underline{0.2293} & \textbf{0.2431}$^*$ \\
            & 15    & 0.1769 & 0.1714  & 0.1272 & 0.1428   & 0.1981 & 0.1368  & 0.2105 & \underline{0.2181} & \textbf{0.2295}$^*$ \\
            & 20    & 0.1415 & 0.1909  & 0.1052 & 0.2129   & 0.1880 & 0.1739  & \underline{0.2358} & 0.1880 & \textbf{0.2758}$^*$ \\ \midrule
\multirow{3}{*}{MCX} 
            & 5     & 0.2131 & 0.2500  & 0.2500 & 0.2501   & \underline{0.2686} & 0.2131  & 0.1164 & 0.2388 & \textbf{0.3194}$^*$ \\
            & 10    & 0.1304 & 0.1846  & 0.2001 & 0.2089   & 0.1718 & \underline{0.2203}  & 0.1342 & 0.2173 & \textbf{0.3508}$^*$ \\
            & 15    & 0.1791 & 0.2343  & 0.1875 & 0.1818   & 0.2328 & 0.2695  & 0.1689 & \underline{0.2933} & \textbf{0.3205}$^*$ \\
            & 20    & 0.2388 & 0.2272  & 0.1600 & 0.2352   & 0.2676 & 0.2631  & 0.1063 & \underline{0.2781} & \textbf{0.3589}$^*$ \\ \midrule
\multirow{3}{*}{ASS09} 
            & 5     & 0.0504 & 0.0771  & 0.0625 & 0.0498   & 0.0971 & 0.0759  & 0.0966 & \underline{0.1245} & \textbf{0.1268}$^*$ \\ 
            & 10    & 0.0916 & 0.0849  & 0.0750 & 0.0696   & \underline{0.1024} & 0.0760   & 0.1018 & 0.0926 & \textbf{0.1174}$^*$ \\
            & 15    & 0.0734 & 0.0482  & 0.0843 & 0.0836   & 0.0931 & \underline{0.1273}  & 0.1183 & 0.0996 & \textbf{0.1281}$^*$ \\
            & 20    & 0.0666 & 0.0675  & 0.1111 & 0.0616   & 0.0769 & 0.1417  & 0.1195 & \underline{0.1434} & \textbf{0.1455}$^*$ \\ \bottomrule
\end{tabular}
\label{tab:comparison}
\end{table*}

\subsubsection{Similarity Agent} \label{sec:Similarity Agent}
While the prerequisite relationships between knowledge concepts provide a basis for recommending a learning path, a learner’s failure to master a key concept may hinder the effective acquisition of subsequent knowledge.
To address this issue, we introduced a similarity agent (S-agent).
The S-agent identifies a set of similar knowledge concepts that are related to the current knowledge concept $c_t$. From this set, the S-agent selects a sub-path that can effectively enhance the learner’s mastery of the current knowledge concept. This sub-path is then appended to the learning path generated by the P-agent.
Please note that, except for its use of a similarity-based knowledge concept structure graph as input, the S-agent operates identically to the P-agent in terms of architecture.

\subsubsection{Difficulty Agent}
To prevent the recommended exercises from being too difficult or too easy, which could lead to a loss of interest during the learning process \cite{MinZD2018, ZhaBL2022}.
We need to find the exercise $e_m$ whose difficulty is closest to the learner's knowledge state.
\begin{equation} \label{equation:difficulty-agent}
     \begin{aligned}
        m=\text{argmin}_{m}\left| \text{difficult}\left( e_{m}^{c_i} \right) -h^{c_i} \right| ,
        \end{aligned}
    \end{equation}
where $m$ represents the exercise ID that best matches the difficulty for the learner, the symbol $\left| \cdot \right|$ represents the absolute value, $\text{difficult}(e^{c_i})$ represents the difficulty of the exercise $e^{c_i}$ corresponding to the KC $c_i$, and $h^{c_i}$ indicates the learner's mastery level of the KC $c_i$.

\subsubsection{Learning Path Construction Mechanism}
In this section, we detail the strategy for selecting initial nodes as well as the communication mechanism among agents.

\looseness=-1
\textit{a) Node Initialization:} Due to the vast number of knowledge concepts, this selection process results in a large search space.
To reduce this complexity, it is essential to identify a preliminary shortest learning path between an initial node and the learning objective before the P-agent performs its selection.
Existing advanced methods, such as adaptive action space learning algorithms, employ the A* algorithm \cite{DucBK2014} to generate the shortest paths while dynamically determining candidate action spaces.
Nonetheless, these methods often fail to identify an effective initial node, which may result in the selection of unsuitable knowledge concepts at the outset and ultimately lead to an ineffective learning path.

\looseness=-1
To address this limitation, we introduce an initial node identification algorithm to ensure the selection of the most appropriate initial node.
Specifically, our approach begins from the target knowledge concept node and iteratively traces back along the prerequisite relationship.
During this backward traversal, we assess whether the prerequisite knowledge concepts of the current node have already been mastered by the learner.
If these prerequisite concepts are mastered, the backward traversal terminates.
Otherwise, we continue tracing back to the node with the longest chain of prerequisite knowledge concepts.
By incorporating this method, the most suitable initial node can be determined before dynamically identifying the candidate action space, effectively optimizing the overall learning path generation.

\looseness=-1
\textit{b) Agent Switching:}
At step $t$, the P-agent evaluates the improvement in the learner’s mastery of prerequisite KC $c_{t-1}$ when selecting the next knowledge concept $c_t$ to recommend.
The evaluation is based on the following:
\begin{equation} \label{equation:S-agent}
     \begin{aligned}
        p_t=h_{t-1}^{c_t}-h_{t-2}^{c_t},
        \end{aligned}
    \end{equation}
where $h_{t-1}^{c_t}$ represents the learner’s knowledge mastery level associated with the knowledge concept $c_t$ learned at step $t-1$, and $h_{t-2}^{c_t}$ represents the knowledge mastery level at step $t-2$.
If the improvement $p_t$ is below a predefined threshold $\tau$, the S-agent is activated.
Once the S-agent completes its selecting process, the P-agent resumes its operation, selecting subsequent KCs based on their prerequisite relationship.
After the P-Agent and S-Agent select the knowledge concepts, we incorporate a difficulty agent to match the selected concepts with exercises that align with the learner's knowledge mastery level.

\section{Experiments}
In this section, we first introduce the experimental setup. Then, through extensive experiments, we demonstrate the superiority of our method based on the experimental results.

\subsection{Datasets}
We conduct our experiments using three publicly available educational datasets to validate the effectiveness of our model: Junyi \footnote{https://pslcdatashop.web.cmu.edu/DatasetInfo?datasetId=1198}, MOOCCubeX \footnote{https://github.com/THU-KEG/MOOCCubeX} - Computer (MCX), and ASSISTments2009 \footnote{https://sites.google.com/site/assistmentsdata/home/2009-2010-assistment-data} (ASS09).
All three educational datasets include learner activity data, which comprises learning records as well as the corresponding knowledge concept names.
Please note that the Junyi dataset provides a comprehensive KC structure graph, illustrating the prerequisite relationships among knowledge concepts. In contrast, the MOOCCubex dataset contains an incomplete KC structure graph, while the ASS09 dataset lacks a KC structure graph altogether. Furthermore, none of the datasets include information on the similarity relationships between KCs.
However, all three datasets contain fields with knowledge concept names, allowing us to construct KC structure graphs for both prerequisite and similarity relationships using our method.
The statistical details of the datasets are shown in Table \ref{Dataset Statistics}.

\begin{table}[ht]
\centering
\caption{Dataset Statistics}
\label{Dataset Statistics}
\begin{tabular}{c|c|c|c}
\hline
\textbf{Dataset} & \textbf{Junyi} & \textbf{MOOCCubeX} & \textbf{ASS09} \\ \hline
\#KCs        & 835            & 443                & 167               \\ 
\#Learners        & 525,061        & 629                & 4,217             \\ 
\#Records         & 21,460,249     & 17,447             & 346,860           \\ 
Positive label rate & 54.38\%       & 70.18\%           & 63.81\%           \\ \hline
\end{tabular}
\vspace{-0.3cm}
\end{table}

\subsection{Compared Methods} \label{CM}
To demonstrate the effectiveness of our method, we compare it with the following baseline methods:
\begin{itemize}[leftmargin=*]
    \item KNN \cite{CovH1967}: The K-Nearest Neighbors algorithm identifies the K closest neighboring values to the target. We use the cosine similarity between learning paths to measure the similarity between two learners and recommend the most similar learning paths.
    \item GRU4Rec \cite{Hid2015}: GRU4Rec is a session-based recommendation model based on Gated Recurrent Units (GRU). It trains an RNN model to capture patterns and information within sequences, predicting the probability of the next exercise appearing.
    \item Actor-Critic \cite{KonT1999}: This method encodes input data using a GRU and then employs a vanilla actor-critic framework for learning path recommendation.
    \item RL-Tutor \cite{KubFM2021}: RL-Tutor is an RL-based adaptive tutoring system that simulates student behavior using DAS3H \cite{DwiKB2018} and recommends learning paths via reinforcement learning.
    \item CSEAL \cite{LiuTL2019}: CSEAL uses DKT \cite{PieBH2015}  to estimate knowledge states and designs a cognitive navigation algorithm to guide the vanilla actor-critic framework in recommending learning paths.
    \item SRC \cite{CheSX2023}: SRC uses a concept-aware encoder to capture the relationships between knowledge concepts, and then a decoder generates learning paths.
    \item GEHRL \cite{LiXY2023}: GEHRL employs a hierarchical reinforcement learning approach, where the lower-level agent filters relevant knowledge concepts for the target, while the higher-level agent is responsible for planning the sequence.
    \item DLPR \cite{ZhaSX2024}: DLPR uses a hierarchical graph neural network to aggregate information about learning tasks and difficulty and then applies a hierarchical reinforcement learning framework to plan knowledge concepts and recommend exercises.
\end{itemize}

\looseness=-1
We evaluate these methods based on the overall improvement of learning objectives after recommendation.
Specifically, in Eq. \eqref{equation:learning score}, $E_{\text{sup}}$ represents the total sum of goals for all learners, while $E_{\text{start}}$ and $E_{\text{end}}$ represent the total sum of initial and final scores for all learners, respectively.

\subsection{Implementation Detail}
We implemented our proposed learning path recommendation framework using PyTorch \cite{PasGM2019} and the simulator code from previous works\cite{LiXY2023, ZhaSX2024}.
In the process of generating KC structures graph, we adjusted the value of the chunk based on the number of KCs in the dataset. The chunk values used for the Junyi, ASS09, and MOOCCubex datasets were $800$, $600$, and $400$, respectively.
In the process of recommending learning path, the first 60\% of exercise logs in the dataset were used to train the environment KT, while the remaining exercise logs were used to train and test our model.
To more accurately evaluate the model's performance, we trained a DKT \cite{PieBH2015} on the entire dataset to predict learners' mastery of target knowledge concepts in Eq. \eqref{equation:learning score}.
Further, similar to previous methods \cite{ZhaSX2024,LiXY2023,LiuTL2019}, we considered a learner's mastery of a KC to be greater than $0.5$, as this implies a probability greater than $0.5$ of correctly answering an exericise related to that KC.
Through testing, we observed that when using our DKT \cite{PieBH2015} for evaluation, the average improvement in learners' knowledge state per exercise was $0.001$.
Thus, in section \ref{sec:Similarity Agent}, we set the threshold $\tau$ to 0.001. We used the Adam optimizer \cite{DieJ2015} for both the policy network and the value network, with a learning rate of 0.001.

\subsection{Overall Performance}
Table \ref{tab:comparison} presents the overall performance of all methods across the three datasets, revealing several key insights.
In comparison with all baseline models, our method outperforms the others on all three datasets.
This demonstrates that leveraging LLM-generated knowledge concept graphs, combined with the integration of similarity relationships, contributes significantly to improving recommendation performance.
Specifically, EDU-GraphRAG effectively mitigates the hallucination problem associated with knowledge concept structure graphs generated by LLMs, which then serve as reliable guidance for learning path recommendation.
Moreover, the incorporation of similarity relationships facilitates smoother learning progression, thereby enhancing the overall performance.

\looseness=-1
Furthermore, most baseline methods exhibit a decline in performance or fail to achieve their best results when the recommendation step reaches 20.
In contrast, KnowLP consistently achieves optimal performance at step=20 across all three datasets.
This further demonstrates that the similarity-aware design effectively smooths the learning process.
\begin{table}[h]
\centering
\caption{Ablation Experiment: ``w/o S'' represents the model without similarity agent, and ``w S'' represents the overall model.}
\begin{tabular}{lcccc cc}
\toprule
\multirow{2}{*}{} & \multicolumn{2}{c}{Junyi} & \multicolumn{2}{c}{MOOCCubeX} & \multicolumn{2}{c}{ASS09}  \\
\cmidrule(lr){2-3} \cmidrule(lr){4-5} \cmidrule(lr){6-7}
 & w S & w/o S & w S & w/o S & w S & w/o S  \\
\midrule
step=5  & \textbf{0.2406} & 0.1932 & \textbf{0.3194} & 0.2148 & \textbf{0.1268} & 0.0737  \\
step=10 & \textbf{0.2431} & 0.2051 & \textbf{0.3508} & 0.1928 & \textbf{0.1174} & 0.1007  \\
step=15 & \textbf{0.2295} & 0.1851 & \textbf{0.3205} & 0.2258 & \textbf{0.1281} & 0.1190  \\
step=15 & \textbf{0.2758} & 0.2232 & \textbf{0.3589} & 0.1517 & \textbf{0.1455} & 0.1264  \\
\bottomrule
\label{tab:Ablation Experiment}
\end{tabular}
\end{table}
\vspace{-0.5cm}
\subsection{Ablation Study}
To validate the effectiveness of the innovative components of our method, we conducted an ablation study. Specifically, we removed the similarity agent module from our method to assess its contribution to the overall performance.
As shown in Table \ref{tab:Ablation Experiment}, when the similarity agent is removed from our method, there is a noticeable performance drop across all three datasets. This highlights the necessity and value of considering the similarity relationships between knowledge concepts in learning path recommendations.
Moreover, with the similarity agent included, our model achieves a substantial performance improvement at step=20.
This further confirms that similarity relationships play a critical role in unlocking the performance potential of longer learning paths.

\begin{figure}[h]
  \centering
  \includegraphics[width=\linewidth]{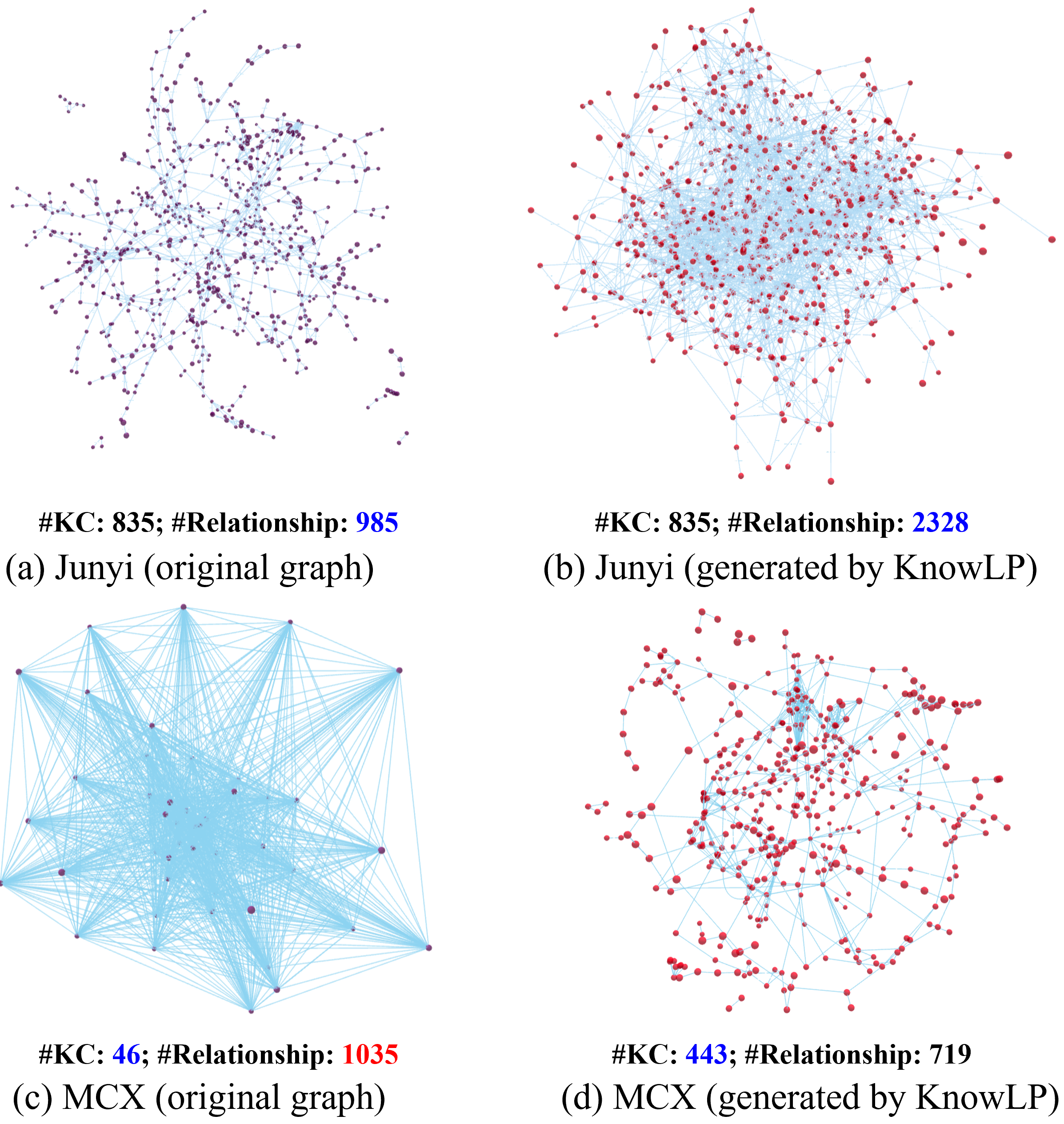}
  \caption{Comparison between the original graph and the graph generated by knowLP. KC denotes the number of knowledge concepts, while Relationship represents the number of prerequisite relationships among these KCs.}
  \vspace{-0.3cm}
  \label{fig:p_graph}
\end{figure}

\begin{figure}[h]
  \centering
  \includegraphics[width=\linewidth]{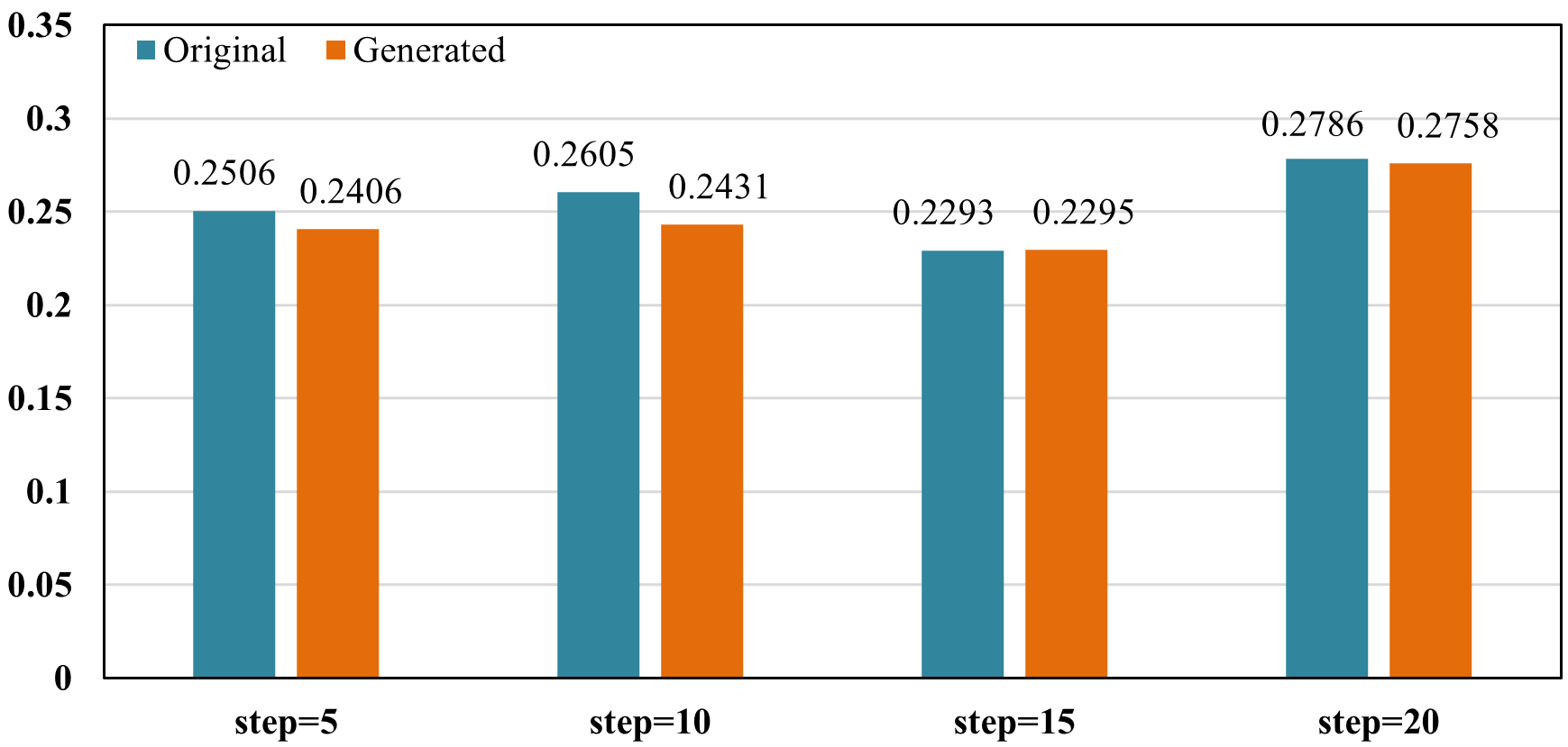}
  \caption{Performance comparison of the original graph and the knowLP-generated graph.}
  \vspace{-0.4cm}
  \label{fig:Knowledge Graph Generation}
\end{figure}

\subsection{Knowledge Graph Generation}
To demonstrate the effectiveness of the KC prerequisite relationship graph generated by KnowLP, we visualized both the original and generated KC structure graphs for the Junyi and MCX datasets. The visualization results of the KC prerequisite relationship graphs are shown in Figure \ref{fig:p_graph}.
It is important to note that since ASS09 does not explicitly provide the KC prerequisite relationship graph, it is not included in the analysis.
We observe that in the Junyi, the number of relationships in the generated KC prerequisite relationship graph is higher than in the original graph, indicating that KnowLP captures more complex prerequisite relationships between KCs.
In the MCX, the original knowledge structure graph, as shown in Figure~\ref{fig:p_graph} (c), covers only a small portion of the 443 KCs in the dataset. Specifically, it includes 1,035 relationships among just 46 KCs, forming a fully connected graph in which every pair of KCs is linked. However, the relations are largely homogeneous, offering little value for learning path generation due to the lack of meaningful prerequisite differentiation. In contrast, the graph generated by KnowLP incorporates all KCs and maintains a more reasonable number of relationships.

\looseness=-1
These observations collectively demonstrate that our method generates more informative and structurally sound knowledge concept prerequisite graphs, addressing key limitations in the original datasets, such as missing or incorrect relationships and incomplete concept coverage.
Moreover, none of the datasets originally contain similarity relationships, further highlighting the value of our approach in enriching the KC structure to support more effective LPR.

\begin{figure}[ht]
  \centering
  \includegraphics[width=\linewidth]{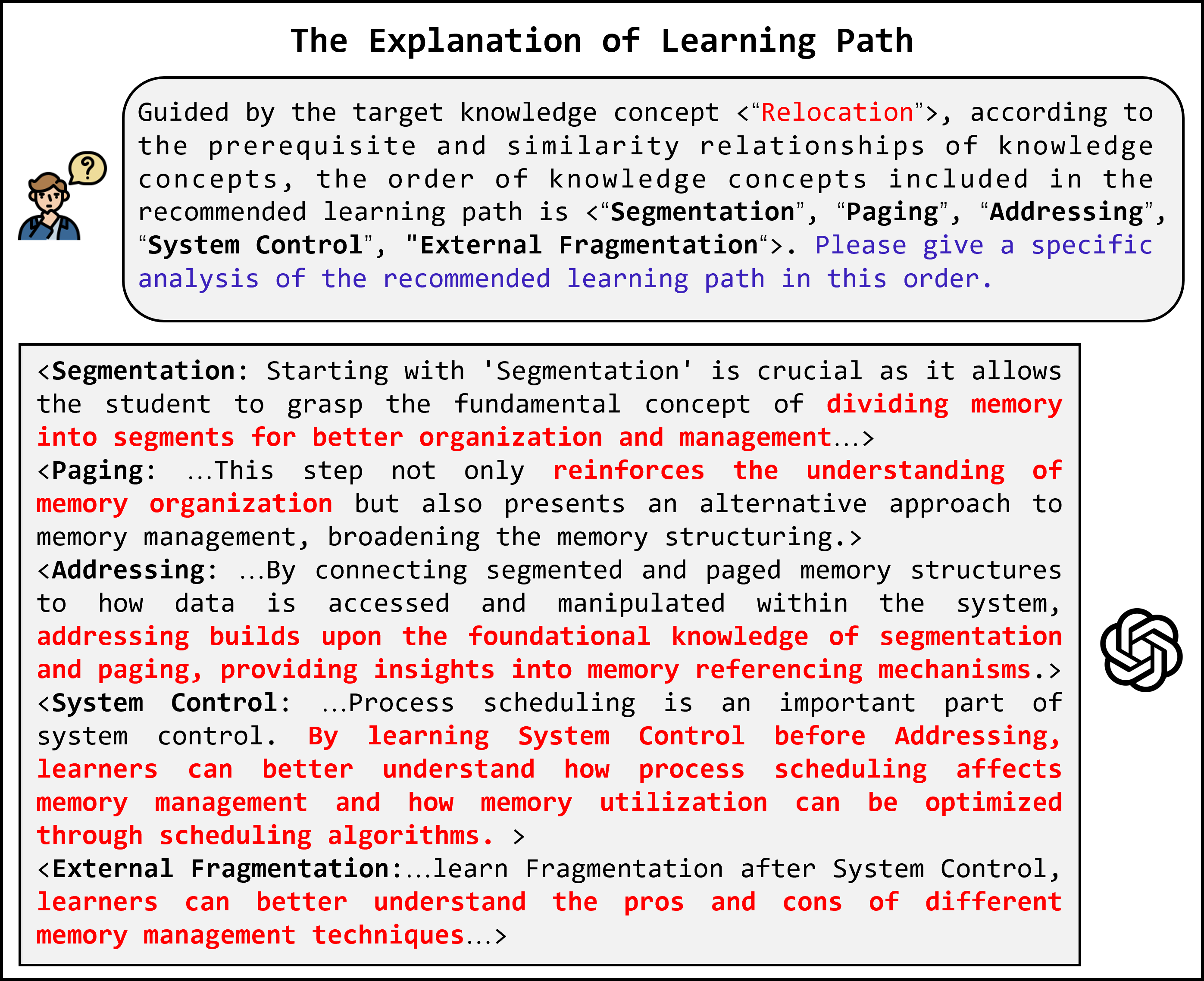}
  \caption{The explanation of learning path.}
  \vspace{-0.5cm}
  \label{fig:Exp_case}
\end{figure}

\begin{figure}[h]
  \centering
  \includegraphics[width=\linewidth]{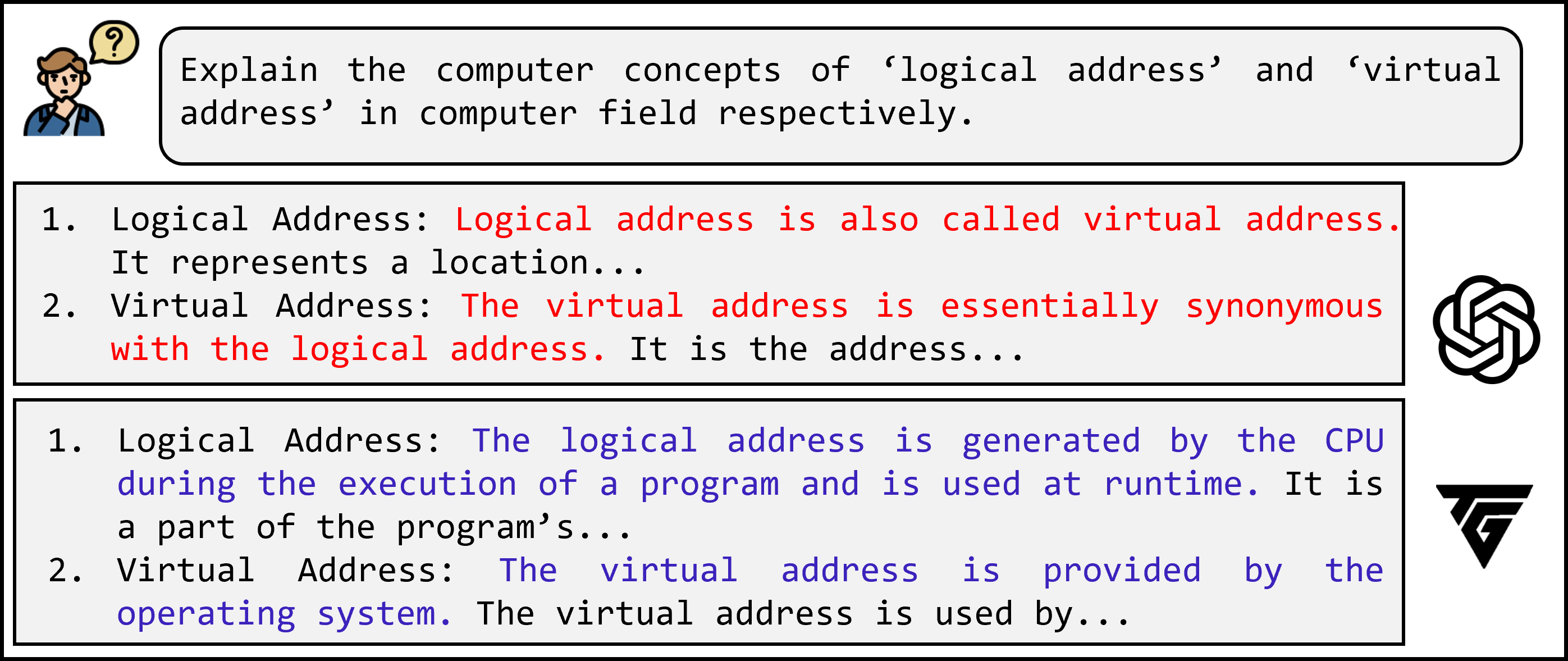}
  \caption{Compare the Results of with/without TextGrad.}
  \vspace{-0.5cm}
  \label{fig:Textgrad_case}
\end{figure}

\subsection{Case Study}
To validate the effectiveness of the innovative components of our method, we conducted an ablation study. Specifically, we removed the similarity agent module from our method to assess its contribution to the overall performance.
As shown in Table \ref{tab:Ablation Experiment}, when the similarity agent is removed from our method, there is a noticeable performance drop across all three datasets. This highlights the necessity and value of considering the similarity relationships between knowledge concepts in learning path recommendations.
Moreover, with the similarity agent included, our model achieves a substantial performance improvement at step=20.
This further confirms that similarity relationships play a critical role in unlocking the performance potential of longer learning paths.

\looseness=-1
In the process of generating the KC structure graph, EDU-Graph RAG also generates a community summary containing key entities, relationships, and claims, which provides useful contextual information for subsequent queries. Taking advantage of this feature of EDU-Graph RAG, we analyzed and explained the recommended learning paths, which improves the interpretability of the recommendations and enhances learners' trust in the recommended content. Figure \ref{fig:Exp_case} shows the analysis and explanation of the recommended paths by EDU-Graph RAG. For each recommended exercise, EDU-Graph RAG is able to explain the reasoning behind the recommendation of the corresponding KC.

\begin{figure}[h]
  \centering
  \includegraphics[width=\linewidth]{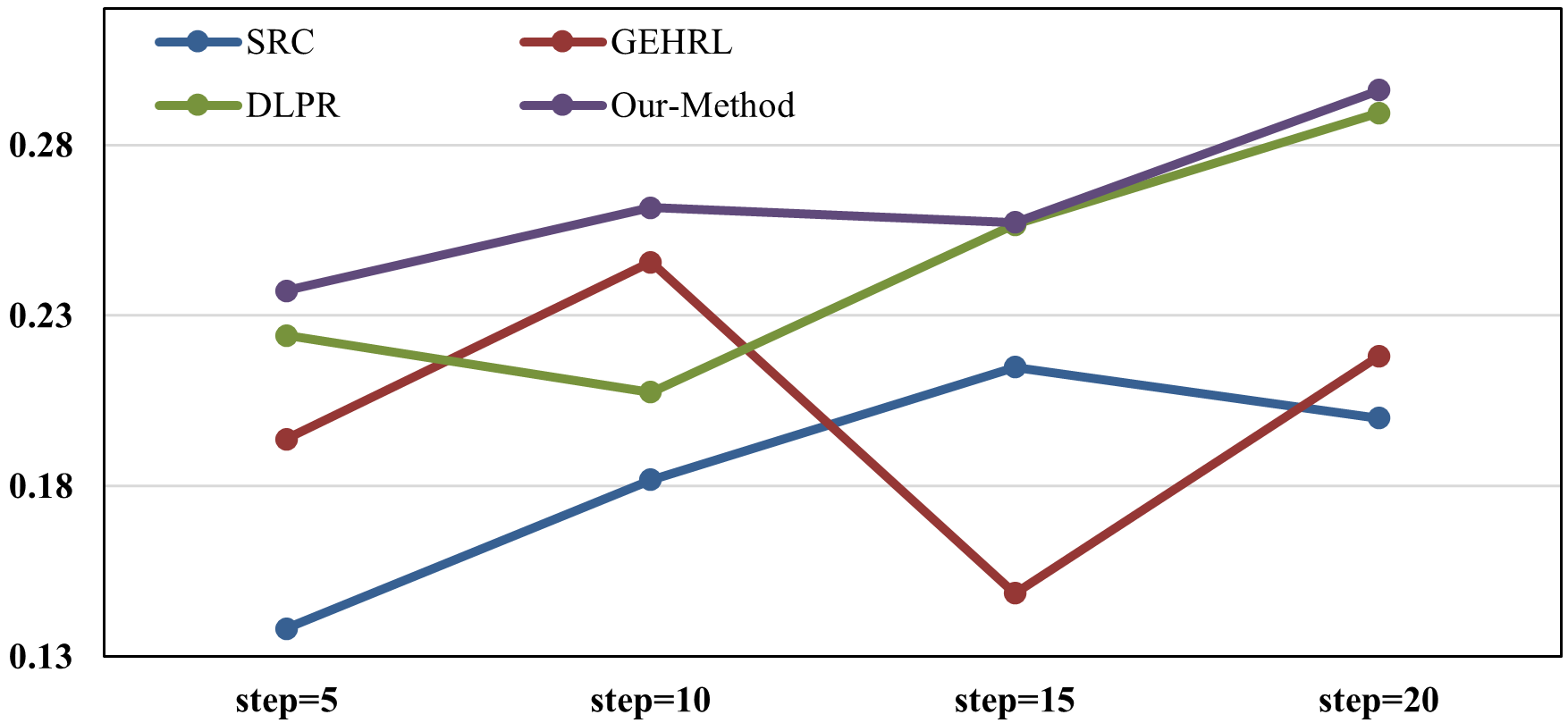}
  \caption{Results of simulation experiment.}
  \vspace{-0.5cm}
  \label{fig:Simulation Experiment}
\end{figure}

\subsection{Simulation Experiment}
Since real-world data only contains static information, which cannot be directly used to analyze the exercise sequences of students not included in the dataset, the effectiveness of our method in real online education scenarios cannot be directly validated. Therefore, we used a Knowledge Evolution-based Simulator (KES), which leverages the DKT model \cite{PieBH2015} to simulate students' exercise behavior on randomly generated exercise sequences. Subsequently, our method performs learning path recommendations based on the student behavior generated by the simulator to validate its effectiveness in real online education scenarios.

\looseness=-1
We constructed the KES-Junyi simulator \cite{LiuTL2019} based on the Junyi dataset. Using simulated initial data, we compared our method with three advanced learning path recommendation methods—SRC, GEHRL, and DLPR—to demonstrate the effectiveness of our approach in online education settings.
The results of the simulation experiments are shown in Figure \ref{fig:Simulation Experiment}. We observe that our method performs better than the other models on the simulated data. This indicates that, when faced with complex and dynamic learner information, our method exhibits strong adaptability, outperforming existing advanced learning path recommendation methods, and is well-suited for application in online education scenarios.

\section{Conclusion}
In this paper, we proposed a KnowLP method to achieve a more effective learning path by considering dual KC structures prerequisite relationships and similarity relationships.
To be specific, we proposed an EDU-GraphRAG module that adaptively generates graphs for knowledge concepts based on diverse educational datasets, enhancing the generalizability of existing learning path recommendation methods.
Then, we developed a DLRL module that alleviates the blocked phenomenon, further improving the performance of learning path recommendations.
Finally, we conducted extensive experiments on three datasets, demonstrating that our method not only achieves state-of-the-art performance but also generates more effective longer learning paths.

\bibliographystyle{ACM-Reference-Format}
\bibliography{main}


\begin{thebibliography}{45}


\ifx \showCODEN    \undefined \def \showCODEN     #1{\unskip}     \fi
\ifx \showDOI      \undefined \def \showDOI       #1{#1}\fi
\ifx \showISBNx    \undefined \def \showISBNx     #1{\unskip}     \fi
\ifx \showISBNxiii \undefined \def \showISBNxiii  #1{\unskip}     \fi
\ifx \showISSN     \undefined \def \showISSN      #1{\unskip}     \fi
\ifx \showLCCN     \undefined \def \showLCCN      #1{\unskip}     \fi
\ifx \shownote     \undefined \def \shownote      #1{#1}          \fi
\ifx \showarticletitle \undefined \def \showarticletitle #1{#1}   \fi
\ifx \showURL      \undefined \def \showURL       {\relax}        \fi
\providecommand\bibfield[2]{#2}
\providecommand\bibinfo[2]{#2}
\providecommand\natexlab[1]{#1}
\providecommand\showeprint[2][]{arXiv:#2}

\bibitem[Cai et~al\mbox{.}(2018)]%
        {CaiXY2018}
\bibfield{author}{\bibinfo{person}{Xiaoyan Cai}, \bibinfo{person}{Junwei Han}, \bibinfo{person}{Wenjie Li}, \bibinfo{person}{Renxian Zhang}, \bibinfo{person}{Shirui Pan}, {and} \bibinfo{person}{Libin Yang}.} \bibinfo{year}{2018}\natexlab{}.
\newblock \showarticletitle{A three-layered mutually reinforced model for personalized citation recommendation}.
\newblock \bibinfo{journal}{\emph{IEEE transactions on neural networks and learning systems}} \bibinfo{volume}{29}, \bibinfo{number}{12} (\bibinfo{year}{2018}), \bibinfo{pages}{6026--6037}.
\newblock


\bibitem[Chen(2009)]%
        {Che2009}
\bibfield{author}{\bibinfo{person}{Chih-Ming Chen}.} \bibinfo{year}{2009}\natexlab{}.
\newblock \showarticletitle{{Ontology-based concept map for planning a personalised learning path}}.
\newblock \bibinfo{journal}{\emph{British Journal of Educational Technology}} \bibinfo{volume}{40}, \bibinfo{number}{6} (\bibinfo{year}{2009}), \bibinfo{pages}{1028--1058}.
\newblock


\bibitem[Chen et~al\mbox{.}(2023)]%
        {CheSX2023}
\bibfield{author}{\bibinfo{person}{Xianyu Chen}, \bibinfo{person}{Jian Shen}, \bibinfo{person}{Wei Xia}, \bibinfo{person}{Jiarui Jin}, \bibinfo{person}{Yakun Song}, \bibinfo{person}{Weinan Zhang}, \bibinfo{person}{Weiwen Liu}, \bibinfo{person}{Menghui Zhu}, \bibinfo{person}{Ruiming Tang}, \bibinfo{person}{Kai Dong}, {et~al\mbox{.}}} \bibinfo{year}{2023}\natexlab{}.
\newblock \showarticletitle{{Set-to-sequence Ranking-based Concept-aware Learning Path Recommendation}}. In \bibinfo{booktitle}{\emph{Proceedings of 2023 AAAI Conference on Artificial Intelligence (AAAI-2023)}}. \bibinfo{pages}{5027--5035}.
\newblock


\bibitem[Cheng et~al\mbox{.}(2025)]%
        {CheXH2025}
\bibfield{author}{\bibinfo{person}{Xinghe Cheng}, \bibinfo{person}{Xufang Zhou}, \bibinfo{person}{Liangda Fang}, \bibinfo{person}{Chaobo He}, \bibinfo{person}{Yuyu Zhou}, \bibinfo{person}{Weiqi Luo}, \bibinfo{person}{Zhiguo Gong}, {and} \bibinfo{person}{Quanlong Guan}.} \bibinfo{year}{2025}\natexlab{}.
\newblock \showarticletitle{NR4DER: Neural Re-ranking for Diversified Exercise Recommendation}. In \bibinfo{booktitle}{\emph{Proceedings of the 48th International ACM SIGIR Conference on Research and Development in Information Retrieval (SIGIR-2025)}}. \bibinfo{pages}{1738–1747}.
\newblock


\bibitem[Chinda(2022)]%
        {Chi2022}
\bibfield{author}{\bibinfo{person}{WOROKWU Chinda}.} \bibinfo{year}{2022}\natexlab{}.
\newblock \showarticletitle{{Effect of Gagne’s learning hierarchy on Chemistry Student achievement in senior secondary school}}.
\newblock \bibinfo{journal}{\emph{Iconic Research and Engineering Journals}} \bibinfo{volume}{5}, \bibinfo{number}{9} (\bibinfo{year}{2022}), \bibinfo{pages}{609--618}.
\newblock


\bibitem[Cover and Hart(1967)]%
        {CovH1967}
\bibfield{author}{\bibinfo{person}{Thomas Cover} {and} \bibinfo{person}{Peter Hart}.} \bibinfo{year}{1967}\natexlab{}.
\newblock \showarticletitle{{Nearest neighbor pattern classification}}.
\newblock \bibinfo{journal}{\emph{IEEE transactions on information theory}} \bibinfo{volume}{13}, \bibinfo{number}{1} (\bibinfo{year}{1967}), \bibinfo{pages}{21--27}.
\newblock


\bibitem[Diederik and Ba(2015)]%
        {DieJ2015}
\bibfield{author}{\bibinfo{person}{P.~Kingma Diederik} {and} \bibinfo{person}{Jimmy~Lei Ba}.} \bibinfo{year}{2015}\natexlab{}.
\newblock \showarticletitle{{Adam: A Method for Stochastic Optimization}}. In \bibinfo{booktitle}{\emph{Proceedings of the 3rd International Conference for Learning Representations (ICLR-2015)}}.
\newblock


\bibitem[Ducho{\v{n}} et~al\mbox{.}(2014)]%
        {DucBK2014}
\bibfield{author}{\bibinfo{person}{Franti{\v{s}}ek Ducho{\v{n}}}, \bibinfo{person}{Andrej Babinec}, \bibinfo{person}{Martin Kajan}, \bibinfo{person}{Peter Be{\v{n}}o}, \bibinfo{person}{Martin Florek}, \bibinfo{person}{Tom{\'a}{\v{s}} Fico}, {and} \bibinfo{person}{Ladislav Juri{\v{s}}ica}.} \bibinfo{year}{2014}\natexlab{}.
\newblock \showarticletitle{{Path planning with modified a star algorithm for a mobile robot}}.
\newblock \bibinfo{journal}{\emph{Procedia engineering}}  \bibinfo{volume}{96} (\bibinfo{year}{2014}), \bibinfo{pages}{59--69}.
\newblock


\bibitem[Dwivedi et~al\mbox{.}(2018)]%
        {DwiKB2018}
\bibfield{author}{\bibinfo{person}{Pragya Dwivedi}, \bibinfo{person}{Vibhor Kant}, {and} \bibinfo{person}{Kamal~K Bharadwaj}.} \bibinfo{year}{2018}\natexlab{}.
\newblock \showarticletitle{{Learning path recommendation based on modified variable length genetic algorithm}}.
\newblock \bibinfo{journal}{\emph{Education and information technologies}}  \bibinfo{volume}{23} (\bibinfo{year}{2018}), \bibinfo{pages}{819--836}.
\newblock


\bibitem[Edge et~al\mbox{.}(2024)]%
        {EdgTC2024}
\bibfield{author}{\bibinfo{person}{Darren Edge}, \bibinfo{person}{Ha Trinh}, \bibinfo{person}{Newman Cheng}, \bibinfo{person}{Joshua Bradley}, \bibinfo{person}{Alex Chao}, \bibinfo{person}{Apurva Mody}, \bibinfo{person}{Steven Truitt}, {and} \bibinfo{person}{Jonathan Larson}.} \bibinfo{year}{2024}\natexlab{}.
\newblock \showarticletitle{{From local to global: A graph rag approach to query-focused summarization}}.
\newblock \bibinfo{journal}{\emph{arXiv preprint arXiv:2404.16130}} (\bibinfo{year}{2024}).
\newblock


\bibitem[Gao et~al\mbox{.}(2023)]%
        {GaoXG2023}
\bibfield{author}{\bibinfo{person}{Yunfan Gao}, \bibinfo{person}{Yun Xiong}, \bibinfo{person}{Xinyu Gao}, \bibinfo{person}{Kangxiang Jia}, \bibinfo{person}{Jinliu Pan}, \bibinfo{person}{Yuxi Bi}, \bibinfo{person}{Yi Dai}, \bibinfo{person}{Jiawei Sun}, {and} \bibinfo{person}{Haofen Wang}.} \bibinfo{year}{2023}\natexlab{}.
\newblock \showarticletitle{{Retrieval-augmented generation for large language models: A survey}}.
\newblock \bibinfo{journal}{\emph{arXiv preprint arXiv:2312.10997}} (\bibinfo{year}{2023}).
\newblock


\bibitem[Goodwin et~al\mbox{.}(2020)]%
        {GooSD2020}
\bibfield{author}{\bibinfo{person}{Travis~R Goodwin}, \bibinfo{person}{Max~E Savery}, {and} \bibinfo{person}{Dina Demner-Fushman}.} \bibinfo{year}{2020}\natexlab{}.
\newblock \showarticletitle{{Flight of the PEGASUS? Comparing Transformers on Few-shot and Zero-shot Multi-document Abstractive Summarization}}. In \bibinfo{booktitle}{\emph{Proceedings of 2020 COLING. international conference on computational linguistics (COLING-2020)}}. NIH Public Access, \bibinfo{pages}{5640}.
\newblock


\bibitem[Guan et~al\mbox{.}(2025)]%
        {GuaQL2025}
\bibfield{author}{\bibinfo{person}{Quanlong Guan}, \bibinfo{person}{Xinghe Cheng}, \bibinfo{person}{Fang Xiao}, \bibinfo{person}{Zhuzhou Li}, \bibinfo{person}{Chaobo He}, \bibinfo{person}{Liangda Fang}, \bibinfo{person}{Guanliang Chen}, \bibinfo{person}{Zhiguo Gong}, {and} \bibinfo{person}{Weiqi Luo}.} \bibinfo{year}{2025}\natexlab{}.
\newblock \showarticletitle{Explainable exercise recommendation with knowledge graph}.
\newblock \bibinfo{journal}{\emph{Neural Networks}}  \bibinfo{volume}{183} (\bibinfo{year}{2025}), \bibinfo{pages}{106954}.
\newblock


\bibitem[Hidasi(2015)]%
        {Hid2015}
\bibfield{author}{\bibinfo{person}{B Hidasi}.} \bibinfo{year}{2015}\natexlab{}.
\newblock \showarticletitle{{Session-based Recommendations with Recurrent Neural Networks}}.
\newblock \bibinfo{journal}{\emph{arXiv preprint arXiv:1511.06939}} (\bibinfo{year}{2015}).
\newblock


\bibitem[Intayoad et~al\mbox{.}(2020)]%
        {IntKT2020}
\bibfield{author}{\bibinfo{person}{Wacharawan Intayoad}, \bibinfo{person}{Chayapol Kamyod}, {and} \bibinfo{person}{Punnarumol Temdee}.} \bibinfo{year}{2020}\natexlab{}.
\newblock \showarticletitle{{Reinforcement learning based on contextual bandits for personalized online learning recommendation systems}}.
\newblock \bibinfo{journal}{\emph{Wireless Personal Communications}} \bibinfo{volume}{115}, \bibinfo{number}{4} (\bibinfo{year}{2020}), \bibinfo{pages}{2917--2932}.
\newblock


\bibitem[Konda and Tsitsiklis(1999)]%
        {KonT1999}
\bibfield{author}{\bibinfo{person}{Vijay Konda} {and} \bibinfo{person}{John Tsitsiklis}.} \bibinfo{year}{1999}\natexlab{}.
\newblock \showarticletitle{{Actor-critic algorithms}}.
\newblock \bibinfo{journal}{\emph{Advances in neural information processing systems}}  \bibinfo{volume}{12} (\bibinfo{year}{1999}).
\newblock


\bibitem[Kubotani et~al\mbox{.}(2021)]%
        {KubFM2021}
\bibfield{author}{\bibinfo{person}{Yoshiki Kubotani}, \bibinfo{person}{Yoshihiro Fukuhara}, {and} \bibinfo{person}{Shigeo Morishima}.} \bibinfo{year}{2021}\natexlab{}.
\newblock \showarticletitle{{Rltutor: Reinforcement learning based adaptive tutoring system by modeling virtual student with fewer interactions}}.
\newblock \bibinfo{journal}{\emph{arXiv preprint arXiv:2108.00268}} (\bibinfo{year}{2021}).
\newblock


\bibitem[Laskar et~al\mbox{.}(2020)]%
        {LasHH2020}
\bibfield{author}{\bibinfo{person}{Md~Tahmid~Rahman Laskar}, \bibinfo{person}{Enamul Hoque}, {and} \bibinfo{person}{Jimmy Huang}.} \bibinfo{year}{2020}\natexlab{}.
\newblock \showarticletitle{{Query Focused Abstractive Summarization via Incorporating Query Relevance and Transfer Learning with Transformer Models}}.
\newblock In \bibinfo{booktitle}{\emph{Proceedings of 2020 Advances in Artificial Intelligence - 33rd Canadian Conference on Artificial Intelligence (Canadian AI-2020)}}. \bibinfo{series}{Lecture Notes in Computer Science}, Vol.~\bibinfo{volume}{12109}. \bibinfo{publisher}{Springer}, \bibinfo{pages}{342--348}.
\newblock


\bibitem[Lewis et~al\mbox{.}(2020)]%
        {LewPP2020}
\bibfield{author}{\bibinfo{person}{Patrick Lewis}, \bibinfo{person}{Ethan Perez}, \bibinfo{person}{Aleksandra Piktus}, \bibinfo{person}{Fabio Petroni}, \bibinfo{person}{Vladimir Karpukhin}, \bibinfo{person}{Naman Goyal}, \bibinfo{person}{Heinrich K{\"u}ttler}, \bibinfo{person}{Mike Lewis}, \bibinfo{person}{Wen-tau Yih}, \bibinfo{person}{Tim Rockt{\"a}schel}, {et~al\mbox{.}}} \bibinfo{year}{2020}\natexlab{}.
\newblock \showarticletitle{{Retrieval-augmented Generation for Knowledge-intensive Nlp Tasks}}. In \bibinfo{booktitle}{\emph{Proceedings of 2020 Advances in Neural Information Processing Systems (NeurIPS-2020)}}. \bibinfo{pages}{9459--9474}.
\newblock


\bibitem[Li et~al\mbox{.}(2024)]%
        {LiXY2024}
\bibfield{author}{\bibinfo{person}{Qingyao Li}, \bibinfo{person}{Wei Xia}, \bibinfo{person}{Li'ang Yin}, \bibinfo{person}{Jiarui Jin}, {and} \bibinfo{person}{Yong Yu}.} \bibinfo{year}{2024}\natexlab{}.
\newblock \showarticletitle{{Privileged Knowledge State Distillation for Reinforcement Learning-based Educational Path Recommendation}}. In \bibinfo{booktitle}{\emph{Proceedings of the 30th ACM SIGKDD Conference on Knowledge Discovery and Data Mining (KDD-2024)}}. \bibinfo{pages}{1621--1630}.
\newblock


\bibitem[Li et~al\mbox{.}(2023)]%
        {LiXY2023}
\bibfield{author}{\bibinfo{person}{Qingyao Li}, \bibinfo{person}{Wei Xia}, \bibinfo{person}{Li'ang Yin}, \bibinfo{person}{Jian Shen}, \bibinfo{person}{Renting Rui}, \bibinfo{person}{Weinan Zhang}, \bibinfo{person}{Xianyu Chen}, \bibinfo{person}{Ruiming Tang}, {and} \bibinfo{person}{Yong Yu}.} \bibinfo{year}{2023}\natexlab{}.
\newblock \showarticletitle{{Graph Enhanced Hierarchical Reinforcement Learning for Goal-oriented Learning Path Recommendation}}. In \bibinfo{booktitle}{\emph{Proceedings of the 32nd ACM International Conference on Information and Knowledge Management (CIKM-2023)}}. \bibinfo{pages}{1318--1327}.
\newblock


\bibitem[Lin et~al\mbox{.}(2013)]%
        {LinYH2013}
\bibfield{author}{\bibinfo{person}{Chun~Fu Lin}, \bibinfo{person}{Yu-chu Yeh}, \bibinfo{person}{Yu~Hsin Hung}, {and} \bibinfo{person}{Ray~I Chang}.} \bibinfo{year}{2013}\natexlab{}.
\newblock \showarticletitle{{Data mining for providing a personalized learning path in creativity: An application of decision trees}}.
\newblock \bibinfo{journal}{\emph{Computers \& Education}}  \bibinfo{volume}{68} (\bibinfo{year}{2013}), \bibinfo{pages}{199--210}.
\newblock


\bibitem[Liu et~al\mbox{.}(2019)]%
        {LiuTL2019}
\bibfield{author}{\bibinfo{person}{Qi Liu}, \bibinfo{person}{Shiwei Tong}, \bibinfo{person}{Chuanren Liu}, \bibinfo{person}{Hongke Zhao}, \bibinfo{person}{Enhong Chen}, \bibinfo{person}{Haiping Ma}, {and} \bibinfo{person}{Shijin Wang}.} \bibinfo{year}{2019}\natexlab{}.
\newblock \showarticletitle{{Exploiting Cognitive Structure for Adaptive Learning}}. In \bibinfo{booktitle}{\emph{Proceedings of the 25th ACM SIGKDD international conference on knowledge discovery \& data mining (KDD-2019)}}. \bibinfo{pages}{627--635}.
\newblock


\bibitem[Minn et~al\mbox{.}(2018)]%
        {MinZD2018}
\bibfield{author}{\bibinfo{person}{Sein Minn}, \bibinfo{person}{Feida Zhu}, {and} \bibinfo{person}{Michel~C Desmarais}.} \bibinfo{year}{2018}\natexlab{}.
\newblock \showarticletitle{{Improving Knowledge Tracing Model by Integrating Problem Difficulty}}. In \bibinfo{booktitle}{\emph{Proceedings of 2018 IEEE International conference on data mining workshops (ICDMW-2018)}}. IEEE, \bibinfo{pages}{1505--1506}.
\newblock


\bibitem[Nabizadeh et~al\mbox{.}(2020)]%
        {NabGG2020}
\bibfield{author}{\bibinfo{person}{Amir~Hossein Nabizadeh}, \bibinfo{person}{Daniel Goncalves}, \bibinfo{person}{Sandra Gama}, \bibinfo{person}{Joaquim Jorge}, {and} \bibinfo{person}{Hamed~N Rafsanjani}.} \bibinfo{year}{2020}\natexlab{}.
\newblock \showarticletitle{{Adaptive learning path recommender approach using auxiliary learning objects}}.
\newblock \bibinfo{journal}{\emph{Computers \& Education}}  \bibinfo{volume}{147} (\bibinfo{year}{2020}), \bibinfo{pages}{103777}.
\newblock


\bibitem[Nabizadeh et~al\mbox{.}(2017)]%
        {NabMP2017}
\bibfield{author}{\bibinfo{person}{Amir~Hossein Nabizadeh}, \bibinfo{person}{Al{\'\i}pio M{\'a}rio~Jorge}, {and} \bibinfo{person}{Jos{\'e} Paulo~Leal}.} \bibinfo{year}{2017}\natexlab{}.
\newblock \showarticletitle{{Rutico: Recommending Successful Learning Paths under Time Constraints}}. In \bibinfo{booktitle}{\emph{Proceedings of the 25th conference on user modeling, adaptation and personalization (UMAP-2017)}}. \bibinfo{pages}{153--158}.
\newblock


\bibitem[Pan et~al\mbox{.}(2024)]%
        {PanSM2024}
\bibfield{author}{\bibinfo{person}{Steven~C Pan}, \bibinfo{person}{Ganeash Selvarajan}, {and} \bibinfo{person}{Chanda~S Murphy}.} \bibinfo{year}{2024}\natexlab{}.
\newblock \showarticletitle{{Interleaved pretesting enhances category learning and classification skills.}}
\newblock \bibinfo{journal}{\emph{Journal of Applied Research in Memory and Cognition}} \bibinfo{volume}{13}, \bibinfo{number}{3} (\bibinfo{year}{2024}), \bibinfo{pages}{393}.
\newblock


\bibitem[Paszke et~al\mbox{.}(2019)]%
        {PasGM2019}
\bibfield{author}{\bibinfo{person}{Adam Paszke}, \bibinfo{person}{Sam Gross}, \bibinfo{person}{Francisco Massa}, \bibinfo{person}{Adam Lerer}, \bibinfo{person}{James Bradbury}, \bibinfo{person}{Gregory Chanan}, \bibinfo{person}{Trevor Killeen}, \bibinfo{person}{Zeming Lin}, \bibinfo{person}{Natalia Gimelshein}, \bibinfo{person}{Luca Antiga}, {et~al\mbox{.}}} \bibinfo{year}{2019}\natexlab{}.
\newblock \showarticletitle{{Pytorch: an Imperative Style, High-performance Deep Learning Library}}. In \bibinfo{booktitle}{\emph{Proceedings of 2019 Advances in Neural Information Processing Systems (NeurIPS-2019)}}. \bibinfo{pages}{8024--8035}.
\newblock


\bibitem[Piech et~al\mbox{.}(2015)]%
        {PieBH2015}
\bibfield{author}{\bibinfo{person}{Chris Piech}, \bibinfo{person}{Jonathan Bassen}, \bibinfo{person}{Jonathan Huang}, \bibinfo{person}{Surya Ganguli}, \bibinfo{person}{Mehran Sahami}, \bibinfo{person}{Leonidas~J Guibas}, {and} \bibinfo{person}{Jascha Sohl-Dickstein}.} \bibinfo{year}{2015}\natexlab{}.
\newblock \showarticletitle{{Deep Knowledge Tracing}}.
\newblock \bibinfo{journal}{\emph{Proceedings of the 28th International Conference on Neural Information Processing Systems (NIPS-2015)}}, \bibinfo{pages}{505--513}.
\newblock


\bibitem[Ram et~al\mbox{.}(2023)]%
        {RamLD2023}
\bibfield{author}{\bibinfo{person}{Ori Ram}, \bibinfo{person}{Yoav Levine}, \bibinfo{person}{Itay Dalmedigos}, \bibinfo{person}{Dor Muhlgay}, \bibinfo{person}{Amnon Shashua}, \bibinfo{person}{Kevin Leyton-Brown}, {and} \bibinfo{person}{Yoav Shoham}.} \bibinfo{year}{2023}\natexlab{}.
\newblock \showarticletitle{{In-context retrieval-augmented language models}}.
\newblock \bibinfo{journal}{\emph{Transactions of the Association for Computational Linguistics}}  \bibinfo{volume}{11} (\bibinfo{year}{2023}), \bibinfo{pages}{1316--1331}.
\newblock


\bibitem[Rohrer(2012)]%
        {Roh2012}
\bibfield{author}{\bibinfo{person}{Doug Rohrer}.} \bibinfo{year}{2012}\natexlab{}.
\newblock \showarticletitle{{Interleaving helps students distinguish among similar concepts}}.
\newblock \bibinfo{journal}{\emph{Educational Psychology Review}}  \bibinfo{volume}{24} (\bibinfo{year}{2012}), \bibinfo{pages}{355--367}.
\newblock


\bibitem[Schulman et~al\mbox{.}(2017)]%
        {SchWD2017}
\bibfield{author}{\bibinfo{person}{John Schulman}, \bibinfo{person}{Filip Wolski}, \bibinfo{person}{Prafulla Dhariwal}, \bibinfo{person}{Alec Radford}, {and} \bibinfo{person}{Oleg Klimov}.} \bibinfo{year}{2017}\natexlab{}.
\newblock \showarticletitle{{Proximal policy optimization algorithms}}.
\newblock \bibinfo{journal}{\emph{arXiv preprint arXiv:1707.06347}} (\bibinfo{year}{2017}).
\newblock


\bibitem[Shen et~al\mbox{.}(2022)]%
        {SheHL2022}
\bibfield{author}{\bibinfo{person}{Shuanghong Shen}, \bibinfo{person}{Zhenya Huang}, \bibinfo{person}{Qi Liu}, \bibinfo{person}{Yu Su}, \bibinfo{person}{Shijin Wang}, {and} \bibinfo{person}{Enhong Chen}.} \bibinfo{year}{2022}\natexlab{}.
\newblock \showarticletitle{{Assessing Student's Dynamic Knowledge State by Exploring the Question Difficulty Effect}}. In \bibinfo{booktitle}{\emph{Proceedings of the 45th international ACM SIGIR conference on research and development in information retrieval (SIGIR-2022)}}. \bibinfo{pages}{427--437}.
\newblock


\bibitem[Shi et~al\mbox{.}(2020)]%
        {ShiWX2020}
\bibfield{author}{\bibinfo{person}{Daqian Shi}, \bibinfo{person}{Ting Wang}, \bibinfo{person}{Hao Xing}, {and} \bibinfo{person}{Hao Xu}.} \bibinfo{year}{2020}\natexlab{}.
\newblock \showarticletitle{{A learning path recommendation model based on a multidimensional knowledge graph framework for e-learning}}.
\newblock \bibinfo{journal}{\emph{Knowledge-Based Systems}}  \bibinfo{volume}{195} (\bibinfo{year}{2020}), \bibinfo{pages}{105618}.
\newblock


\bibitem[Wang et~al\mbox{.}(2023)]%
        {WanLY2023}
\bibfield{author}{\bibinfo{person}{Hangyu Wang}, \bibinfo{person}{Ting Long}, \bibinfo{person}{Liang Yin}, \bibinfo{person}{Weinan Zhang}, \bibinfo{person}{Wei Xia}, \bibinfo{person}{Qichen Hong}, \bibinfo{person}{Dingyin Xia}, \bibinfo{person}{Ruiming Tang}, {and} \bibinfo{person}{Yong Yu}.} \bibinfo{year}{2023}\natexlab{}.
\newblock \showarticletitle{{GMOCAT: A Graph-Enhanced Multi-Objective Method for Computerized Adaptive Testing}}. In \bibinfo{booktitle}{\emph{Proceedings of the 29th ACM SIGKDD Conference on Knowledge Discovery and Data Mining (KDD-2023)}}. \bibinfo{pages}{2279--2289}.
\newblock


\bibitem[Wang et~al\mbox{.}(2024)]%
        {WanJP2024}
\bibfield{author}{\bibinfo{person}{Jiapu Wang}, \bibinfo{person}{Sun Kai}, \bibinfo{person}{Linhao Luo}, \bibinfo{person}{Wei Wei}, \bibinfo{person}{Yongli Hu}, \bibinfo{person}{Alan Wee-Chung Liew}, \bibinfo{person}{Shirui Pan}, {and} \bibinfo{person}{Baocai Yin}.} \bibinfo{year}{2024}\natexlab{}.
\newblock \showarticletitle{Large language models-guided dynamic adaptation for temporal knowledge graph reasoning}.
\newblock \bibinfo{journal}{\emph{Advances in Neural Information Processing Systems}}  \bibinfo{volume}{37} (\bibinfo{year}{2024}), \bibinfo{pages}{8384--8410}.
\newblock


\bibitem[Xu et~al\mbox{.}(2012)]%
        {XuWC2012}
\bibfield{author}{\bibinfo{person}{Dihua Xu}, \bibinfo{person}{Zhijian Wang}, \bibinfo{person}{Kejia Chen}, {and} \bibinfo{person}{Weidong Huang}.} \bibinfo{year}{2012}\natexlab{}.
\newblock \showarticletitle{{Personalized Learning Path Recommender Based on User Profile Using Social Tags}}. In \bibinfo{booktitle}{\emph{Proceedings of the 5th International Symposium on Computational Intelligence and Design (ISCID-2012)}}, Vol.~\bibinfo{volume}{1}. IEEE, \bibinfo{pages}{511--514}.
\newblock


\bibitem[Yang et~al\mbox{.}(2023)]%
        {YanYP2023}
\bibfield{author}{\bibinfo{person}{Xiaocheng Yang}, \bibinfo{person}{Mingyu Yan}, \bibinfo{person}{Shirui Pan}, \bibinfo{person}{Xiaochun Ye}, {and} \bibinfo{person}{Dongrui Fan}.} \bibinfo{year}{2023}\natexlab{}.
\newblock \showarticletitle{{Simple and Efficient Heterogeneous Graph Neural Network}}. In \bibinfo{booktitle}{\emph{Proceedings of 2023 AAAI conference on artificial intelligence (AAAI-2023)}}. \bibinfo{pages}{10816--10824}.
\newblock


\bibitem[Yuksekgonul et~al\mbox{.}(2024)]%
        {YukBB2024}
\bibfield{author}{\bibinfo{person}{Mert Yuksekgonul}, \bibinfo{person}{Federico Bianchi}, \bibinfo{person}{Joseph Boen}, \bibinfo{person}{Sheng Liu}, \bibinfo{person}{Zhi Huang}, \bibinfo{person}{Carlos Guestrin}, {and} \bibinfo{person}{James Zou}.} \bibinfo{year}{2024}\natexlab{}.
\newblock \showarticletitle{{TextGrad: Automatic" Differentiation" via Text}}.
\newblock \bibinfo{journal}{\emph{arXiv preprint arXiv:2406.07496}} (\bibinfo{year}{2024}).
\newblock


\bibitem[Yuksekgonul et~al\mbox{.}(2025)]%
        {YukBB2025}
\bibfield{author}{\bibinfo{person}{Mert Yuksekgonul}, \bibinfo{person}{Federico Bianchi}, \bibinfo{person}{Joseph Boen}, \bibinfo{person}{Sheng Liu}, \bibinfo{person}{Pan Lu}, \bibinfo{person}{Zhi Huang}, \bibinfo{person}{Carlos Guestrin}, {and} \bibinfo{person}{James Zou}.} \bibinfo{year}{2025}\natexlab{}.
\newblock \showarticletitle{Optimizing generative AI by backpropagating language model feedback}.
\newblock \bibinfo{journal}{\emph{Nature}}  \bibinfo{volume}{639} (\bibinfo{year}{2025}), \bibinfo{pages}{609--616}.
\newblock


\bibitem[Zhang et~al\mbox{.}(2022)]%
        {ZhaBL2022}
\bibfield{author}{\bibinfo{person}{Haotian Zhang}, \bibinfo{person}{Chenyang Bu}, \bibinfo{person}{Fei Liu}, \bibinfo{person}{Shuochen Liu}, \bibinfo{person}{Yuhong Zhang}, {and} \bibinfo{person}{Xuegang Hu}.} \bibinfo{year}{2022}\natexlab{}.
\newblock \showarticletitle{{APGKT: Exploiting Associative Path on Skills Graph for Knowledge Tracing}}.
\newblock In \bibinfo{booktitle}{\emph{Proceedings of 2022 Pacific Rim International Conference on Artificial Intelligence (PRICAI-2022)}}. \bibinfo{series}{Lecture Notes in Computer Science}, Vol.~\bibinfo{volume}{13629}. \bibinfo{publisher}{Springer}, \bibinfo{pages}{353--365}.
\newblock


\bibitem[Zhang et~al\mbox{.}(2024)]%
        {ZhaSX2024}
\bibfield{author}{\bibinfo{person}{Haotian Zhang}, \bibinfo{person}{Shuanghong Shen}, \bibinfo{person}{Bihan Xu}, \bibinfo{person}{Zhenya Huang}, \bibinfo{person}{Jinze Wu}, \bibinfo{person}{Jing Sha}, {and} \bibinfo{person}{Shijin Wang}.} \bibinfo{year}{2024}\natexlab{}.
\newblock \showarticletitle{{Item-Difficulty-Aware Learning Path Recommendation: From a Real Walking Perspective}}. In \bibinfo{booktitle}{\emph{Proceedings of the 30th ACM SIGKDD Conference on Knowledge Discovery and Data Mining (KDD-2024)}}. \bibinfo{pages}{4167--4178}.
\newblock


\bibitem[Zhang et~al\mbox{.}(2023)]%
        {ZhaZW2023}
\bibfield{author}{\bibinfo{person}{Lei Zhang}, \bibinfo{person}{Wuji Zhang}, \bibinfo{person}{Likang Wu}, \bibinfo{person}{Ming He}, {and} \bibinfo{person}{Hongke Zhao}.} \bibinfo{year}{2023}\natexlab{}.
\newblock \showarticletitle{{SHGCN: Socially enhanced heterogeneous graph convolutional network for multi-behavior prediction}}.
\newblock \bibinfo{journal}{\emph{ACM Transactions on the Web}} \bibinfo{volume}{18}, \bibinfo{number}{1} (\bibinfo{year}{2023}), \bibinfo{pages}{1--27}.
\newblock


\bibitem[Zhang et~al\mbox{.}(2021)]%
        {ZhaLZ2021}
\bibfield{author}{\bibinfo{person}{Qian Zhang}, \bibinfo{person}{Jie Lu}, {and} \bibinfo{person}{Guangquan Zhang}.} \bibinfo{year}{2021}\natexlab{}.
\newblock \showarticletitle{{Recommender Systems in E-learning}}.
\newblock \bibinfo{journal}{\emph{Journal of Smart Environments and Green Computing}} \bibinfo{volume}{1}, \bibinfo{number}{2} (\bibinfo{year}{2021}), \bibinfo{pages}{76--89}.
\newblock


\bibitem[Zheng et~al\mbox{.}(2023)]%
        {ZheCP2023}
\bibfield{author}{\bibinfo{person}{Chuanpan Zheng}, \bibinfo{person}{Xiaoliang Fan}, \bibinfo{person}{Shirui Pan}, \bibinfo{person}{Haibing Jin}, \bibinfo{person}{Zhaopeng Peng}, \bibinfo{person}{Zonghan Wu}, \bibinfo{person}{Cheng Wang}, {and} \bibinfo{person}{S~Yu Philip}.} \bibinfo{year}{2023}\natexlab{}.
\newblock \showarticletitle{Spatio-temporal joint graph convolutional networks for traffic forecasting}.
\newblock \bibinfo{journal}{\emph{IEEE Transactions on Knowledge and Data Engineering}} \bibinfo{volume}{36}, \bibinfo{number}{1} (\bibinfo{year}{2023}), \bibinfo{pages}{372--385}.
\newblock


\end{thebibliography}


\end{document}